\documentstyle[12pt,epsfig,glas]{article}
\newlength{\dinwidth}
\newlength{\dinmargin}
\input epsf
\epsfverbosetrue

\newcommand{\sleq}{\raisebox{-0.5mm}{$\stackrel{<}{\scriptstyle{\sim}}$}}

\def\Journal#1#2#3#4{{#1} {\bf #2}, #3 (#4)}

\def\AIP{{\em AIP Conf. Proc.}}
\def\NPB{{\em Nucl. Phys.} B}
\def\PLB{{\em Phys. Lett.}  B}
\def\EPC{{\em Euro. Phys.}  C}
\def\PRL{{\em Phys. Rev. Lett.}}
\def\PRD{{\em Phys. Rev.} D}
\def\PR{{\em Phys. Rep.}}
\def\ZPC{{\em Z. Phys.} C}

\def\MPL{{\em Mod. Phys. Lett.} A}

\def\st{\scriptstyle}

\def\be{\begin{equation}}
\def\ee{\end{equation}}
\def\bea{\begin{eqnarray}}
\def\eea{\end{eqnarray}}

\def\lapproxeq{\lower .7ex\hbox{$\;\stackrel{\textstyle <}{\sim}\;$}}
\def\gapproxeq{\lower .7ex\hbox{$\;\stackrel{\textstyle >}{\sim}\;$}}
\def\btop{\mathchar"1339}
\def\bmid{\mathchar"133D}
\def\bbot{\mathchar"133B}

\newcommand{\pom}{I\!\!P}
\newcommand{\reg}{I\!\!R}

\newcommand{\Apom}{A_{\pom}}
\newcommand{\Areg}{A_{\reg}}

\newcommand{\alphapom}{\alpha_{_{I\!\!P}}}
\newcommand{\alphareg}{\alpha_{_{I\!\!R}}}

\newcommand{\gsim}{\raisebox{-0.5mm}{$\stackrel{>}{\scriptstyle{\sim}}$}}

\begin{document}
\begin{titlepage}{GLAS-PPE/98--07}{16$^{\underline{\rm{th}}}$ December 1998}
\title{Structure Functions}

\author{Anthony T. Doyle
}


\begin{abstract}
The latest structure function results, 
as presented at the ICHEP98 conference, are
reviewed. A brief introduction to the formalism and the status of 
global analyses of parton distributions is given.
The review focuses on three experimental areas: 
fixed-target results and their constraints on the parton densities at high $x$;
spin structure and spin parton densities as well as the status of the 
associated sum rules; HERA results on the dynamics of $F_2$ at low $(x,Q^2)$, 
charm and $F_L$ as well as the measurement and interpretation of the high-$Q^2$ 
neutral and charged current cross-sections.

\vspace{0.5cm}
\centerline{\em Plenary talk presented at the XXIX ICHEP98 Conference,}
\centerline{\em Vancouver, July 1998.}
\vspace{0.5cm}
\centerline{\em Slides are available from}
\centerline{\em http://www-zeus.desy.de/conferences98/\#ichep98}
\end{abstract}
\newpage
\end{titlepage}


\section{Introduction - Formalism and Road Maps}
The differential cross-section $l(k)N(p)\rightarrow l(k')X(p')$
for a lepton ($e$, $\mu$) 
with four-momentum $k$ scattering off a nucleon with 
four-momentum $p$ can be expressed as 
\begin{eqnarray}
\tiny
\frac{d^2\sigma (l^\pm N)}{dx\,dQ^2} & = & \frac{2 \pi \alpha^2}{x Q^4} \nonumber
\\
&& \cdot [ Y_+ F_2(x, Q^2) \mp Y_- xF_3(x, Q^2) - y^2 F_L(x, Q^2)]  \nonumber
\end{eqnarray}
\noindent
where 
$Q^2$ is the four-momentum transfer squared, 
$x = Q^2/2p\cdot q$ is the Bjorken scaling variable,
$y = p\cdot q/p\cdot k$ is the inelasticity variable and
$ Y_{\pm} = 1 \pm (1-y)^2$.
The contribution from $F_2$ dominates the cross-section. 
The contribution 
from $F_L$ is a QCD correction which is important only at large $y$
and that from $xF_3$ is negligible for $Q^2 << M_Z^2$.
To investigate sensitivity to $F_L$ at large $y$ or $xF_3$ at large $Q^2$,
the reduced cross-section 
$\tilde{\sigma} \equiv \frac{xQ^4}{2\pi\alpha^2}\frac{1}{Y_+}
\frac{d^2\sigma}{dxdQ^2}$
is adopted.
In the Quark Parton Model (or in the DIS scheme of NLO QCD) 
and for $Q^2 << M_Z^2$ $F_2/x$ is 
the charge-weighted sum of the quark densities 
$$ F_2(x, Q^2) = x \sum_i e_i^2 \cdot \Sigma(x,Q^2) $$
where $\Sigma(x,Q^2) = \sum_i [q_i(x,Q^2) + \bar{q}_i(x,Q^2)]$
is the {\em singlet} summed quark and anti-quark distributions.
Similarly, the charged current cross-section
$e^+(e^-)N \rightarrow \nu(\bar{\nu})X$ 
at HERA can be expressed as
\begin{eqnarray}
\tiny
\frac{d^2\sigma^{CC}(l^\pm N)}{dx\,dQ^2} & 
= & \frac{G_F^2}{4 \pi x}
\left(\frac{M_W^2}{Q^2+M_W^2}\right)^2 \nonumber
\\
&& \cdot [ Y_+ F_2^{CC}(x, Q^2) \mp Y_- xF_3^{CC}(x, Q^2)].  \nonumber
\end{eqnarray}
\noindent
For fixed-target $\nu(\bar{\nu})N\rightarrow \mu^+(\mu^-)X$ 
experiments, $Q^2 << M_W^2$, and in the QPM 
$$ xF_3^{CC}(x, Q^2) = x \sum_i e_i^2 \cdot q_{NS}(x,Q^2)$$
where $q_{NS}(x,Q^2) = \sum_i [q_i(x,Q^2) - \bar{q}_i(x,Q^2)]$
is the {\em non-singlet} difference of these distributions.

\begin{figure}
\center
\epsfig{file=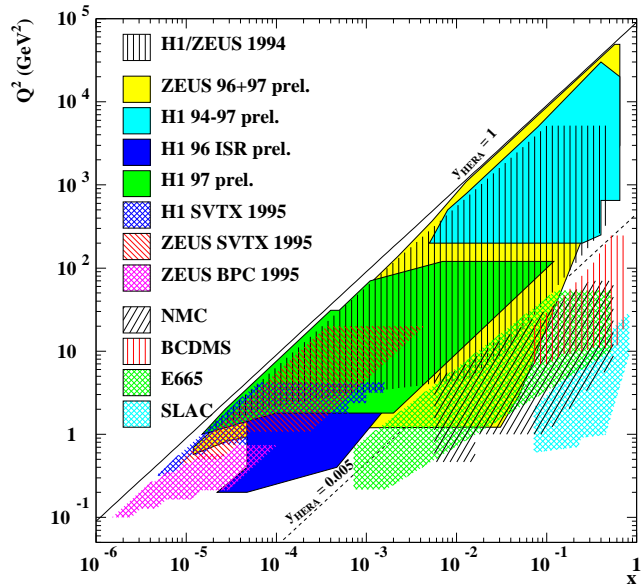,
bbllx=60pt,bblly=150,bburx=570,bbury=650,width=8cm}
\caption{Measured regions of $F_2$ in the $(x,Q^2)$ kinematic plane.
The nominal acceptance region of the HERA measurements corresponds to 
$y_{\rm {HERA}} > 0.005$. 
The fixed-target experimental data occupies the region of high $x$ 
at low $Q^2$.
\label{kine}}
\end{figure}

In Fig.~\ref{kine}, the kinematic plane covered by the $F_2(x,Q^2)$ 
measurements is shown, including the new preliminary datasets from H1 
and ZEUS which are seen to extend to:
low $y$ ($y_{{\rm HERA}} \sim 0.005$) providing overlap
with the fixed-target experiments; 
very low $x$ ($x \sleq 10^{-5}$) at low $Q^2$ exploring the
transition region from soft to hard physics;
high $y$ ($y \rightarrow 0.82$) giving 
sensitivity to $F_L$;
high $x\rightarrow 0.65$ probing sensitivity to 
electroweak effects in $F_2$ and $xF_3$ as well as 
constraining the valence quarks at large $Q^2$. 
The fixed-target experiments NMC, BCDMS, E665 and SLAC experiments 
have provided final measurements at higher $x$ and lower $Q^2$. New 
information was presented at ICHEP98 from CCFR, E866 and the Tevatron which
also constrain the medium-high $x$ partons.

\begin{figure}
\center
\epsfig{file=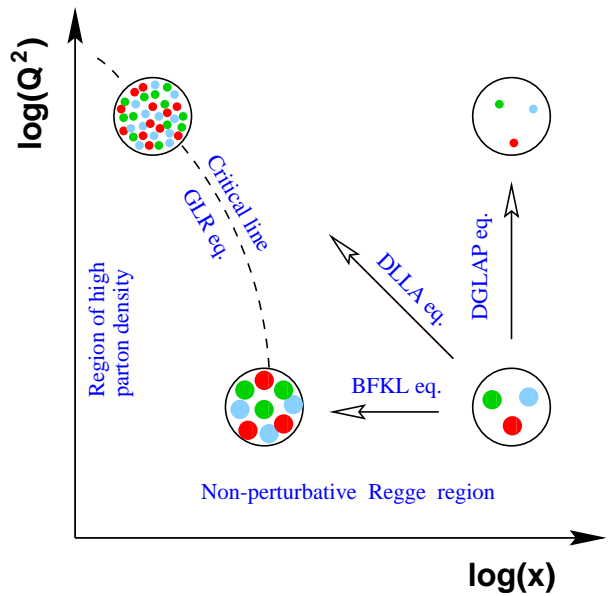,width=8cm}
\caption{Schematic representation of parton densities and the
theoretical evolution directions in the $(x,Q^2)$ kinematic plane.
\label{evol}}
\end{figure}

Theoretically, the directions in $(x,Q^2)$ 
can be mapped out according to the dominant 
dynamical effects, as illustrated in Fig.~\ref{evol}.
Given a phenomenological input as a function of $x$,
the parton distributions are evolved to different physical scales ($Q^2$)
via the DGLAP evolution equations.
The alternative BFKL approach is to attempt to calculate the $x$ dependence
directly from pQCD, where the running of the effective coupling constant
is neglected to leading order. 
BFKL predicts an $x^{-\lambda}$ dependence of $F_2$ at small $x$.
The BFKL equation has recently been calculated 
to NLO.~\cite{fadin} These corrections are numerically 
very large 
in the experimentally accessible low ($x,Q^2$) range,
resulting in
$\lambda_{LO} \sim 0.5$ being reduced to $\lambda_{NLO} \sim 0.1$.
The understanding of the NLO corrections to the BFKL equation is therefore
an active area of study.~\cite{jeff}
In the DLLA (`double leading log approximation'), 
non-leading $\ln(Q^2) \ln(1/x)$ terms can also be evaluated, but a method 
which reliably maps the complete region of $(x,Q^2)$ in terms of
pQCD is still not known.
The 
region of high parton density may be reached at very 
low $x$, where these approaches are not strictly valid.
The expectation from the GLR equation~\cite{GLR} 
is that the region where the partons
overlap is accessible at slightly higher $x$ for decreasing $Q^2$.
However, $Q^2$ should also be sufficiently large that higher twist
and non-perturbative effects parameterised in terms of Regge exchanges 
can be neglected. 

In the DGLAP approach, the non-singlet contribution evolves as
$$\frac{\partial q_{NS}(x,Q^2)}{\partial t} =  {\alpha_s(Q^2)  \over 2 \pi} 
P_{qq}^{NS} \otimes q_{NS}(x',Q^2)$$ 
where $t = \ln (Q^2/\Lambda^2)$ and 
the $P_{ij}$'s represent the NLO DGLAP splitting probabilities for 
radiating a parton with momentum fraction $x$ from a parton with higher
momentum $x'$.
Quantities such as $xF_3$ provide a measure of $\alpha_s(Q^2)$ which
is insensitive to the a priori unknown gluon distribution.
Similarly,
the singlet quark and gluon densities are coupled via 
$$
{\partial \over \partial t} 
\left(\begin{array}{c} \Sigma(x,Q^2)
\\ g(x,Q^2)
\end{array}\right) 
= {\alpha_s(Q^2)
  \over 2 \pi} \left[\begin{array}{cc}
    P_{qq} & P_{qg}  \\
    P_{gq} & P_{gg}
\end{array}\right] \otimes
\left(\begin{array}{c}\Sigma(x',Q^2) \\ g(x',Q^2) \end{array}\right)
$$
and quantities such as $F_2$ provide input for
$\Sigma(x,Q^2)$ as well as coupled knowledge of 
$\alpha_s(Q^2)$ and the gluon, $g(x,Q^2)$. 


At the starting scale, $Q_{\rm o}^2$, the light valence quarks ($q_v$) and the 
sea of quark and anti-quarks ($\bar{q}$) 
as well as the gluon ($g$) are attributed a given functional form. For 
example in the MRST parameterisations  
$$f_i(x,Q_{\rm o}^2) = A_i x^{\delta_i} (1-x)^{\eta_i}(1+\epsilon_i\sqrt{x} + \gamma_i x) $$
where some parameters are set to 0 or fixed by sum rules
and differences of $u$ and $d$ quarks may additionally be constrained.
The heavy quark (sea) contributions are calculated explicitly at NLO
and their uncertainty is typically determined by the range allowed for the 
effective mass of the heavy quark. 
The measured structure functions are then described by the convolutions of the
light quark densities with the appropriate NLO matrix elements.

The outlined procedure defines the structure of a nucleon in terms of its
constituent quarks and gluons.
However, 
guided by the new datasets and analyses which were presented to this 
conference, it should be noted that
the following assumptions are made:
\begin{itemize}
\item 
{$\alpha_s(Q^2) \ln(Q^2)$ (DGLAP) terms are large compared to 
$\alpha_s(Q^2) \ln(1/x)$ (BFKL) terms in the perturbative splitting functions;
\footnote{
The study of inclusive quantities such as $F_2$ at small $x$ 
are presently unable 
to distinguish these BFKL terms. The status of forward jet production searches 
which enhance sensitivity to these effects is reviewed by 
J. Huston.~\cite{joey}}}
\item 
{higher-twist (HT) contributions (suppressed 
by factors of $1/Q^2$) are negligibly small;
\footnote{Progress in determining the size of these higher twist and 
other hadronisation power corrections via infra-red renormalons is reviewed
by Y. Dokshitzer.~\cite{yuri}}}
\item 
{Nuclear binding effects are treated as small corrections or ignored in
analyses of deuteron data.
\footnote{A discussion of diffractive final
states in D.I.S including nuclear effects 
is included in the review by M. Erdmann.~\cite{martin}}}
\end{itemize}

In Table~\ref{tab:mrst}, the experimental datasets considered in the 
MRST analysis are listed along with the underlying physics process and 
the parton behaviour which is being probed.
The experiments denoted by $\dagger$ correspond to final measurements    
reported at ICHEP98. Those denoted by $\star$ correspond to new
preliminary measurements reported at this conference.
The latest global fits of MRST~\cite{MRST}, GRV98~\cite{GRV}
and CTEQ4~\cite{CTEQ4} are used here to compare with the data.

\begin{table*}[htb]
\caption{Processes studied in global parton distribution fits.
$\star$ indicates new preliminary data presented at ICHEP98, included in this 
review. $\dagger$ indicates final data presented at ICHEP98.
$^*$ indicates data used in the MRST fits.
(Courtesy of A. Martin.)
\label{tab:mrst}}
\vspace{0.2cm}
\begin{center}
\begin{tabular}{|l|l|l|}    \hline                                                                                        
{\bf Process/}     &     {\bf Leading order}   & {\bf Parton behaviour probed}\\                                                                                       
{\bf Experiment}   &  {\bf subprocess}         &                           \\                                                                                       
\hline                                                                                       
&\hfill \raisebox{-0.5ex}[0.5ex][0.5ex]{$\btop$}&                      \\                                                                                       
{\bf DIS} $\mbox{\boldmath $(\mu N \rightarrow \mu X)$}$ &  $\gamma^*q                                                                                       
\rightarrow q$ \hfill {\arrayrulewidth=1pt\vline}\hspace*{4pt}&  \\                                                                                       
$F^{\mu p}_2,F^{\mu d}_2,F^{\mu n}_2/F^{\mu p}_2$                                                                                       
& \hfill {\arrayrulewidth=1pt\vline}\hspace*{4pt}&                                                                                         
Four structure                                                                                       
functions $\rightarrow$  \\                                                                                       
(SLAC, BCDMS,& \hfill {\arrayrulewidth=1pt\vline}\hspace*{4pt}& \hspace*{1cm}                                                                                        
$u + \bar{u}$  \\                                                                                       
NMC, E665)$^*$& \hfill {\arrayrulewidth=1pt\vline}\hspace*{4pt}&                                                                                       
\hspace*{1cm} $d + \bar{d}$   \\                                                                                       
      &\hfill $\bmid$ & \hspace*{1cm}  $\bar{u} + \bar{d}$  \\                                                                                      
{\bf DIS} $\mbox{\boldmath $(\nu N \rightarrow \mu X)$}$ & $W^*q \rightarrow                                                                                       
q^{\prime}$  \hfill {\arrayrulewidth=1pt\vline}\hspace*{4pt}& \hspace*{1cm}                                                                                         
$s$ (assumed = $\bar{s}$),\\                                                                                       
$F^{\nu N}_2,xF^{\nu N}_3$    &\hfill 
{\arrayrulewidth=1pt\vline}\hspace*{4pt}&                                                                                      
but only                                                                                       
$\int xg(x,Q_{\rm o}^2)dx \simeq 0.35$ \\                                                                                       
(CCFR)$^*$ {\hfill$\dagger$}    &\hfill 
{\arrayrulewidth=1pt\vline}\hspace*{4pt}&
and $\int(\bar{d}-\bar{u})dx\simeq 0.1$  \\ 
                                                                                    
                &\hfill \raisebox{0.5ex}[1.5ex][1.5ex]{$\bbot$} &         \\                                                                                       
{\bf DIS (small $x$)}    &   $\gamma^* (Z^*)q \rightarrow q$   &   $\lambda$     \\                                                                                       
$F^{ep}_2$ (H1, ZEUS)$^*$   {\hfill$\star$}    &                  & $(x\bar{q} \sim x^{-\lambda_S},                                                                                      
\ xg \sim  x^{-\lambda_g})$    \\[2mm]                                                                                      
{\bf DIS ($\mbox{\boldmath $F_L$}$)}    &   $\gamma^* g \rightarrow q\bar{q}$                     
&   $g$     \\                                                                                       
NMC, HERA   {\hfill$\star$}    &                  &     \\[2mm]                                 
$\mbox{\boldmath $\ell N \rightarrow c \bar{c} X$}$ &                                                                                      
  $\gamma^*  c\rightarrow  c$    & $c$ \\                                                                                       
 $F_2^{c}$  (EMC; H1, ZEUS)$^*$  {\hfill$\star$}              &     &     $  \quad (x \gapproxeq 0.01;\                                                                                       
x \lapproxeq 0.01 )$    \\[2mm]                                                                                      
$\mbox{\boldmath $\nu N \rightarrow \mu^+\mu^-X$}$ &  $W^* s \rightarrow                                                                                        
c$    & $s \approx \frac{1}{4} (\bar{u} + \bar{d}) $ \\                                                                                       
(CCFR)$^*$            &   $\;\;\;\;\;\;\;\;\;\;\;\;\;\hookrightarrow \mu^+$     &         \\[2mm]                                                                                      
$\mbox{\boldmath $p N \rightarrow \gamma X$}$  &  $qg \rightarrow \gamma q$  &                                                                                      
$g$ at $x \simeq 2 p_T^\gamma/\sqrt{s} \rightarrow$ \\                                                                                       
(WA70$^*$, UA6, E706, \ldots)    &      &  \hspace*{.5cm}   $x \approx 0.2 - 0.6  $            
\\[2mm]                                                                                      
$\mbox{\boldmath $pN \rightarrow \mu^+\mu^- X$}$   &  $q\bar{q} \rightarrow                                                                                       
\gamma^*$  &  $\bar{q} = ...(1-x)^{\eta_S}$ \\                                                                                       
(E605, E772)$^*$          &                     &                             \\[2mm]                                                                                      
$\mbox{\boldmath $pp, pn \rightarrow \mu^+\mu^- X$}$ & $u\bar{u},d\bar{d}                                                                                       
\rightarrow \gamma^*$   & $\bar{u} - \bar{d}~~(0.04 \lapproxeq x \lapproxeq 0.3)$                    
\\                                                                                       
(E866, NA51)$^*$     {\hfill$\dagger$}            &  $u\bar{d},d\bar{u} \rightarrow \gamma^*$   &     \\[2mm]                                                                                     
$\mbox{\boldmath $ep, en \rightarrow e \pi X$}$    &  $\gamma^* q \rightarrow q$         
with   &   $\bar{u} - \bar{d}~~(0.04 \lapproxeq x  \lapproxeq 0.2)$     \\                                                                                       
(HERMES)  {\hfill$\dagger$}     & $q = u, d, \bar{u}, \bar{d}$ &     \\[2mm]                                
$\mbox{\boldmath $p\bar{p} \rightarrow WX(ZX)$}$    &  $ud \rightarrow W$                                                                                       
&  $u,d$ at $x \simeq M_W/\sqrt{s} \rightarrow$  \\                                                                                       
(UA1, UA2; CDF, D0)          &   & \hspace*{.5cm}  $x \approx 0.13;~0.05$ \\[2mm]                              
$\;\;\mbox{\boldmath $\rightarrow \ell^{\pm}$}$ {\bf asym} (CDF)$^*$   
{\hfill$\dagger$}            &         &                                                                                       
slope of $u/d$ at $x \approx 0.05 - 0.1$  \\[2mm]                                                                                      
$\mbox{\boldmath $p\bar{p} \rightarrow t\bar{t} X$}$ & $q\bar{q}, gg \rightarrow          
t\bar{t}$ & $q, g$ at $x \gapproxeq 2m_t/\sqrt{s} \simeq 0.2$ \\         
(CDF, D0) &                       & \\[2mm]         
$\mbox{\boldmath $p\bar{p} \rightarrow ${\bf jet}$\, + \, X$}$    &                                                                                       
 $gg,qg,qq\rightarrow 2j$  &  $q,g$                                                                                       
at $x \simeq 2 E_T/\sqrt{s} \rightarrow$  \\                                                                                       
(CDF, D0)         &   & \hspace*{.5cm}  $x \approx 0.05 -  0.5$ \\ \hline                                                                                      
\end{tabular}
\end{center}
\end{table*}

\begin{figure}
\center
\epsfig{file=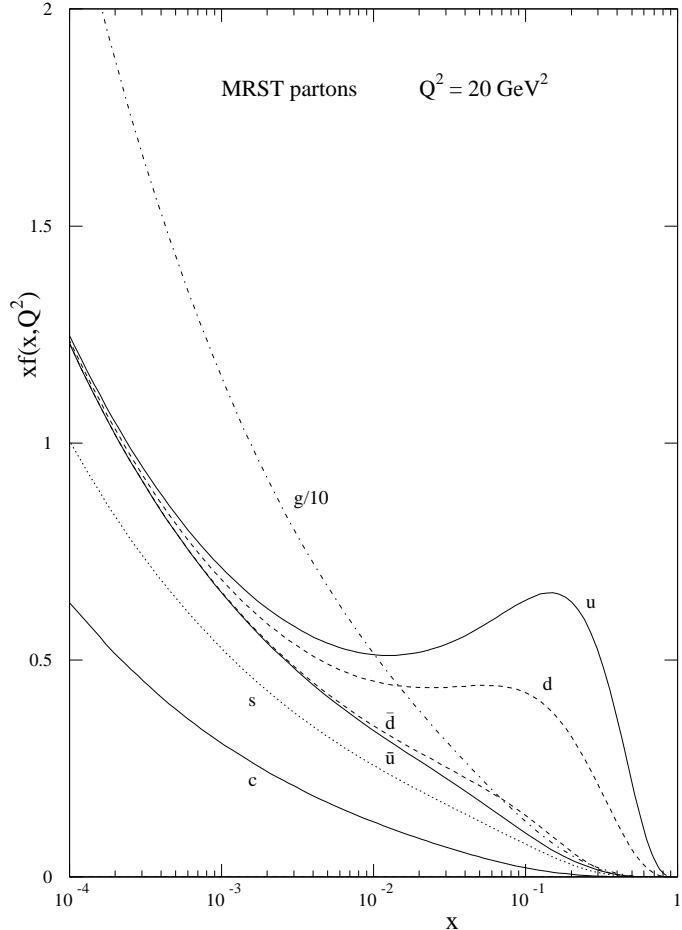,
width=9cm}
\caption{MRST parton distributions $xf(x,Q^2)$ at $Q^2 = 20$~GeV$^2$.
\label{mrs1}}
\end{figure}

The output MRST parton distributions evolved to $Q^2 = 20$~GeV$^2$ are shown in
Fig.~\ref{mrs1}.
At this $Q^2$, 
the expected fractions of the total momentum carried by 
the valence quarks is 
25\%($u_v$) and 10\%($d_v$). 
6\%($2\bar{u}$), 8\%($2\bar{d}$), 
5\%($2s$) and 2\%($2c$) 
is carried by the $q\bar{q}$ symmeteric sea
and 44\%($g$) is carried by the gluons.

Incorporating new datasets and theoretical understanding 
improves the precision with which individual
parton densities are known.
In particular, uncertainties in 
$d/u$ at high $x$,
$\bar{d}/\bar{u}$ at intermediate $x$,
$s$ at all $x$ values,
$g$ at low $x$ and
$c$ at low $x$, 
are discussed in the context of the 
presentations made at ICHEP98.

\section{Fixed-Target Results}
\subsection{Determination of $\alpha_s$}
\begin{figure}
\center
\epsfig{file=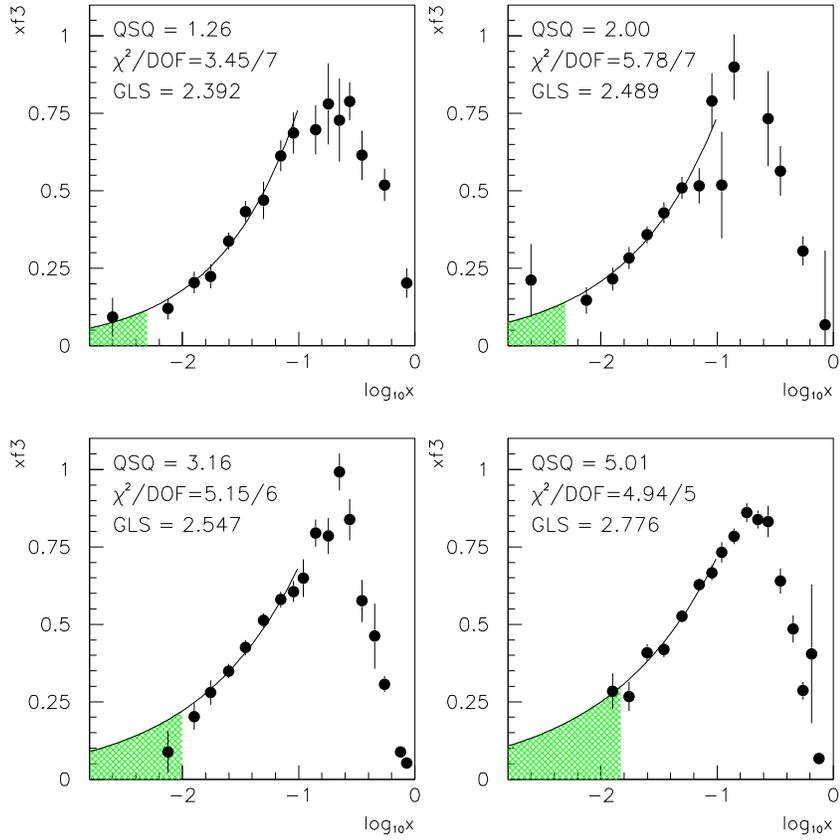,width=12cm}
\caption{CCFR analysis of $xF_3$ versus $log_{10}(x)$ for four ranges of 
$Q^2$.
The curve indicates a power law ($Ax^B$) fit applied for 
$x<0.1$ used to determine the integral in the unmeasured shaded region.
\label{glsr}}
\end{figure}

\noindent{\em Gross-Llewellyn Smith (GLS) Sum Rule:}
The GLS sum rule expresses the fact that there are 
three valence quarks in the nucleon, subject to QCD corrections
$$\tiny \int_0^1 xF_3(x,Q^2) \frac{dx}{x} = 
 3(1-\frac{\alpha_s}{\pi} - a_2(\frac{\alpha_s}{\pi})^2
 - a_3(\frac{\alpha_s}{\pi})^3) - 
 \frac{{\Delta HT}}{Q^2}.$$
This is a fundamental prediction of QCD which, 
as a non-singlet quantity, is independent of the gluon distribution.
The sum rule has been calculated to 
$\mathcal{O}$$(\alpha_s^3)$
and estimated to $\mathcal{O}$$(\alpha_s^4)$.~\cite{larin,kataev}
The CCFR collaboration have now published their final results,~\cite{kim}
incorporating earlier neutrino measurements (WA25, WA59, SKAT,
FNAL-E180 and BEBC-Gargamelle) to determine the GLS sum rule.
In Fig.~\ref{glsr}, the sum rule is proportional to the area under the 
data in four regions of $Q^2$. 
The analysis of the world data enables $\alpha_s(3$~GeV$^2)$ to be determined
at NNLO ($\mathcal{O}$$(\alpha_s^3)$) accuracy and
evolved to $M_Z^2$ as
$$\tiny \alpha_s(M_Z^2) = 0.114 ^{+0.005}_{-0.006}(stat.) 
^{+0.007}_{-0.009}(sys.) ^{+0.004}_{-0.005}(theory).$$
The largest contribution to the systematic error is the uncertainty
on the ratio of the total neutrino and anti-neutrino cross-sections, 
$\sigma_{\bar{\nu}}/\sigma_\nu = 0.499 \pm 0.007$
which determines the overall normalisation of $xF_3$.
This uncertainty will be improved with the NuTeV 
tagged $\nu$-$\bar{\nu}$ beamline.
The nuclear target corrections to the GLS sum rule 
are predicted to be small.~\cite{kulagin}
The largest theory uncertainty is that associated with the higher twist
contribution, $\Delta HT = 0.15 \pm 0.15$~GeV$^2$. 
Here, a renormalon approach, where chains of vacuum polarisation 
bubbles on a gluon propagator lead to a prediction of the higher 
twist contribution in the perturbative expansion, predicts a small 
correction, $\Delta HT < 0.02$~GeV$^2$.
Other models predict significantly larger HT corrections and
the uncertainty encompasses this range.
The renormalon 
approach is now rather successful in describing a range of power 
corrections to hadronic final state variables~\cite{yuri} and 
leads to a central value of $\alpha_s(M_Z^2) = 0.118$.
The most recent analysis of the $xF_3$ data determines the same central
value with a similar theoretical uncertainty.~\cite{kataev2}

\begin{figure}
\center
\epsfig{file=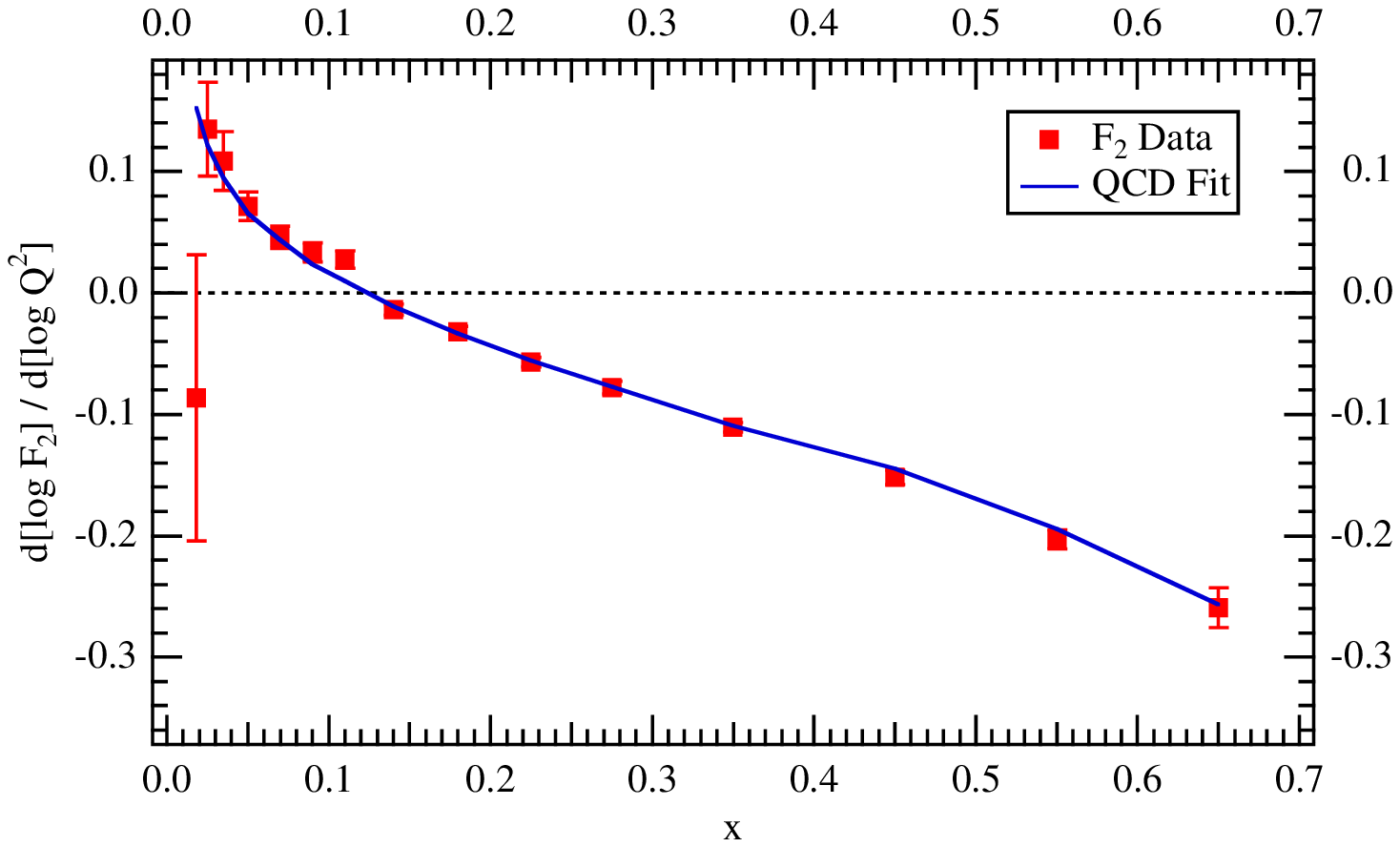,width=12cm}
\epsfig{file=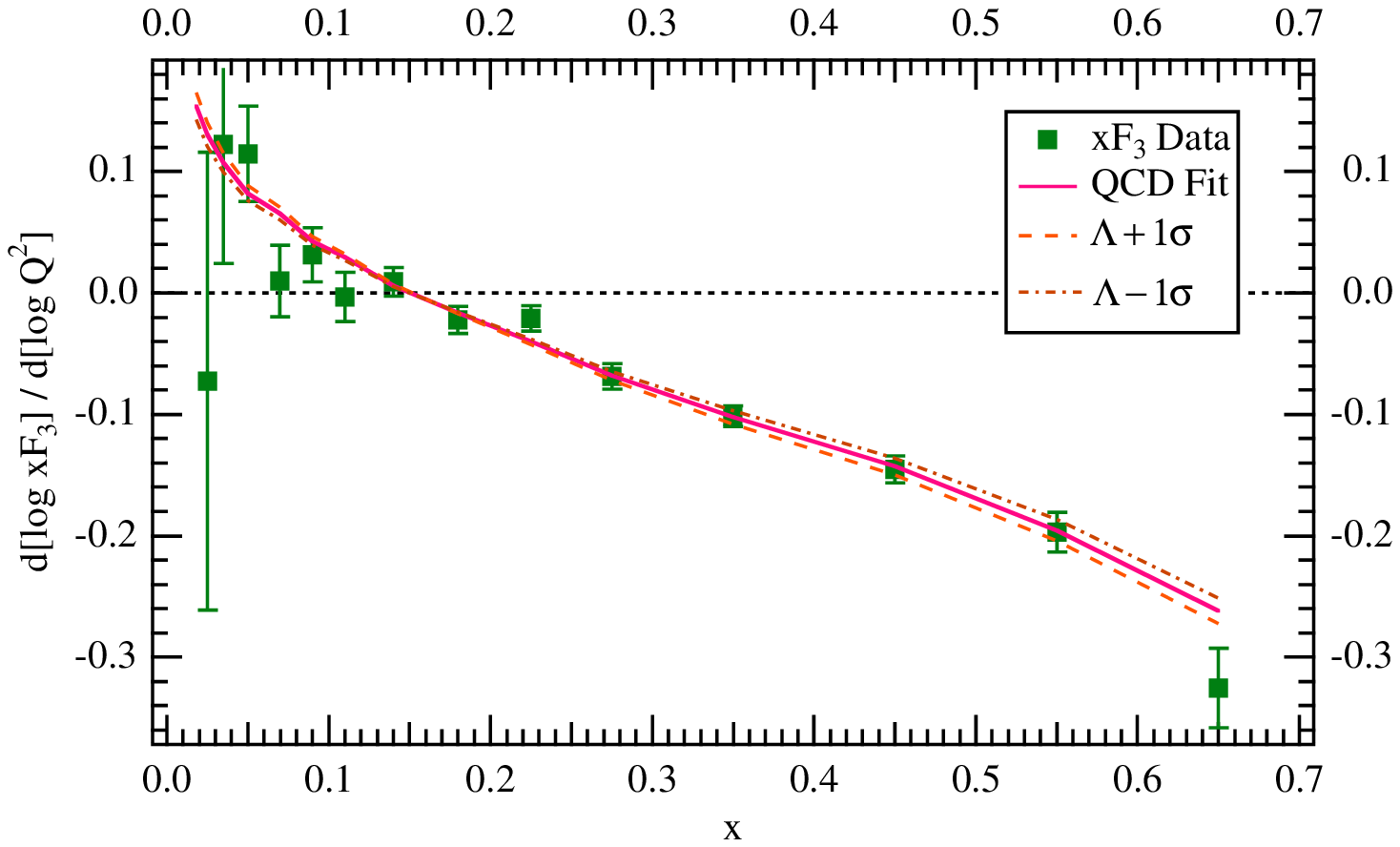,width=12cm}
\caption{CCFR results on the scaling violations of $F_2$ (upper plot)
and $xF_3$ (lower plot) compared to a NLO QCD fit.
\label{hxsv}}
\end{figure}

\noindent{\em Scaling Violations at Large $x$:}
The most accurate method to determine $\alpha_s(M_Z^2)$ from the CCFR data
remains the measurement of the scaling violations of the structure 
functions, $F_2$ and $xF_3$, using a NLO QCD fit.~\cite{seligman}
The scaling violation slopes
$d(\log F_2)/d\log Q^2$
and 
$d(\log xF_3)/d\log Q^2$
are shown in 
Fig.~\ref{hxsv}. From $F_2$, the high-$x$ gluon is also constrained to 
be 
$$xg(x, Q_{\rm o}^2=5{\rm~{GeV}}^2) = (2.22\pm0.34) \times (1-x)^{4.65\pm0.68}$$
in agreement with global analyses using prompt photon data. 
From the combined results on $F_2$ and $xF_3$  
$$\alpha_s(M_Z^2) = 0.119\pm0.002(exp.)\pm0.001(HT)\pm0.004(scale)$$
which represents one of the most precise
determinations of this quantity.
This improved measurement 
is higher than the earlier CCFR value due to the use of a
new energy calibration. It is also 
higher than the SLAC/BCDMS analysis value of 
$\tiny \alpha_s(M_Z^2) = 0.113\pm0.003(exp.)\pm0.004(theory)$.~\cite{virchaux} 
Here it is noted that there is a small but 
statistically significant discrepancy
between the SLAC and BCDMS data 
(see the $x \geq 0.45$ $F_2^p$ points in Fig.~\ref{hi-x}), 
which can be resolved if the 
correlated systematics of the BCDMS data at low $y$ are taken into account.
The quoted central value of the combined analysis, however, 
does not take these systematics into account.~\cite{eisele}
There is therefore little evidence for any discrepancy with respect to
the world average value $\alpha_s^{PDG}(M_Z^2) = 0.119\pm0.002$.~\cite{PDG}

Improved statistics will enable the NuTeV collaboration to determine 
$\alpha_s$ decoupled from the gluon using the $xF_3$ data alone.
Further progress in the $F_2$ analysis requires the calculation of 
NNLO terms in order to reduce 
the renormalisation and factorisation scale uncertainties.
In this way the overall uncertainty on  $\alpha_s(M_Z^2)$ can be reduced 
to $\pm0.002$.
In this area, there is still an outstanding 20\% discrepancy between the 
CCFR $F_2^\nu$ and NMC $F_2^\mu$ data in the region 
of $x \sim 10^{-2}$ which is discussed in the context of the strange quark sea 
and the ZEUS preliminary data later. However, this effect is negligible in the 
determination of $\alpha_s$, which depends mainly upon the high-$x$ data
as seen in Fig.~\ref{hxsv}.

\subsection{Nucleon Structure}
{\em $d/u$ at large $x$:}
Valence parton distributions at high $x$
received attention at this 
conference provoked by new data and 
corrections of the $d/u$ ratio from an analysis
by U.~Yang and A. Bodek.~\cite{yang}
Extractions of $F_2^n$ from NMC and SLAC deuteron data have previously
accounted for Fermi motion but not nuclear binding effects.
A physically appealing model by Frankfurt and Strikman~\cite{frankfurt} 
assumes that binding effects in the deuteron and heavier nuclear targets 
scale with the nuclear density. 
This nuclear binding correction is about 4\% at $x=0.7$ for fixed-target
deuteron experiments, 
which is parameterised as an additional term $\delta(d/u)= 
(0.10 \pm 0.01) x(1+x)$ added to the MRS(R2) PDF. 
This is sufficient to increase the $d$ distribution significantly at high 
$Q^2 = 10^4$~GeV$^2$ by about 40\% at $x=0.5$, due to DGLAP evolution
of the partons.
The modification gives an improved fit to the NMC deuteron data.
It is compelling in that this simple modification now improves 
the description of high-$x$ CDHSW $\nu p / \bar{\nu} p$ data (not shown)
as well as the new data on the $W$ asymmetry from CDF (probing 
intermediate $x \simeq M_W/\sqrt{s} \simeq 0.05$
values) and the charged 
current cross-sections from ZEUS (see later), data which are free from 
nuclear binding effects. 

\begin{figure}
\center
\hspace*{-0.5cm}
\epsfig{file=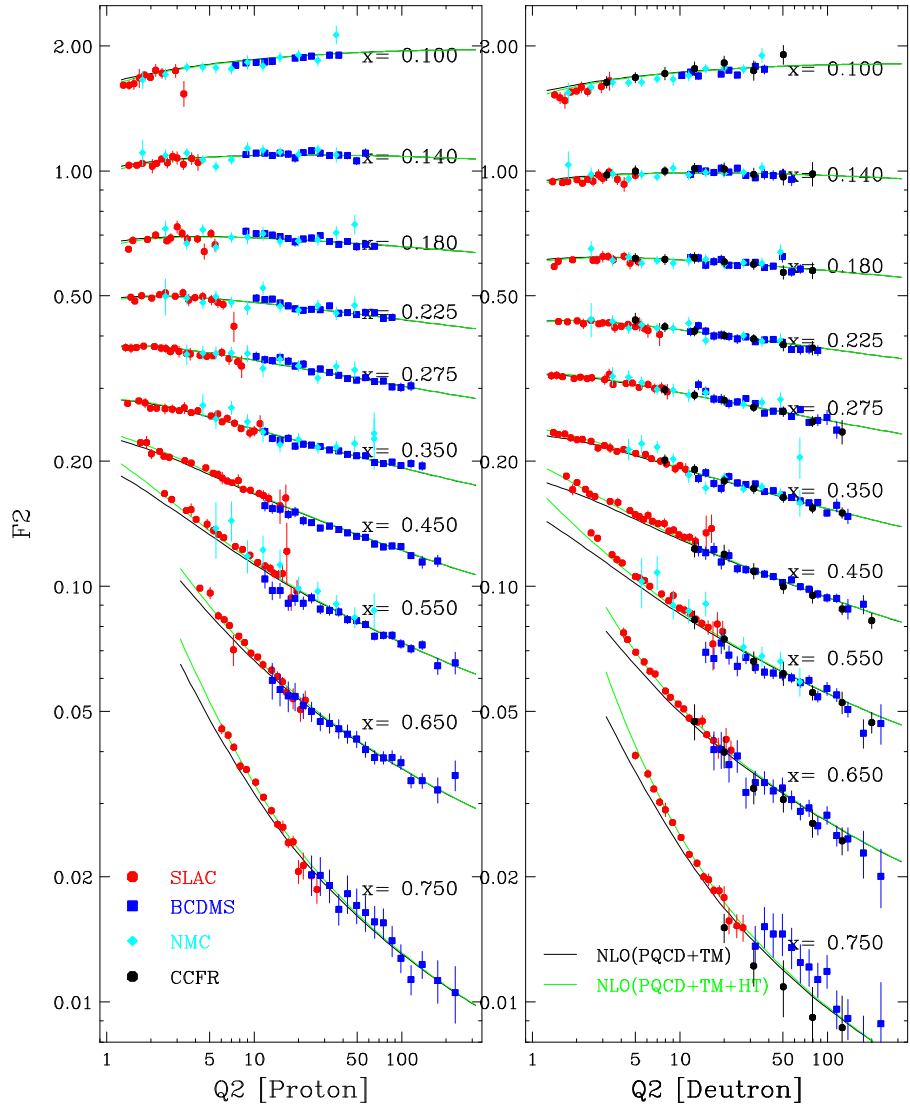,
bbllx=60pt,bblly=90,bburx=506,bbury=640,clip=,width=12cm}
\caption{Comparison of high-$x$ $F_2^p$ and $F_2^d$ data versus $Q^2$
fitted with and without higher twist (HT) corrections. 
The error bars correspond to statistical and systematic errors
added in quadrature.
\label{hi-x}}
\end{figure}

The world $F_2^p$ and $F_2^d$ fixed-target data at high $x$ is plotted 
in Fig.~\ref{hi-x}. Here the NLO pQCD calculation incorporates target 
mass (TM) effects determined using Georgi-Politzer scaling where the scaling
variable $x$ is replaced by 
$\xi = 2x/(1+\sqrt{1+4M^{2}x^{2}/Q^{2}})$
as well as the correction for nuclear binding effects for the deuteron data.
The description of the low $Q^2$ data is significantly improved 
($\chi^2/DoF = 1577/1045$) if higher twist (HT) corrections are incorporated 
as indicated by the fit denoted NLO(pQCD+TM+HT).~\cite{yang}
A good description is also obtained for the very high $x$ SLAC DIS data 
(not shown) up to
$x=0.9$, although there is some residual $Q^2$ dependence in the resonance
region ($0.9 < x <1$).
The renormalon 
approach~\cite{webber} therefore 
successfully predicts the $x$ dependence of the higher twist terms
although the normalisation constant is half of the previously estimated value.
This renormalon approach also successfully describes the $Q^2$ dependence
of $R =\sigma_L/\sigma_T$ (not shown).
In a NNLO fit this higher twist 
contribution tends to zero which suggests that the non-singlet higher 
twist contribution (relevant for $xF_3$) is also small, as discussed in
relation to the GLS sum rule. In addition, new measurements of $R$, covering
the range of $0.015<x<0.55$ from CCFR were presented which are in agreement 
with those from EMC, BCDMS and SLAC and extend the kinematic range up to 
$Q^2 \sim 100$~GeV$^2$.~\cite{jae}

\begin{figure}
\center
\epsfig{file=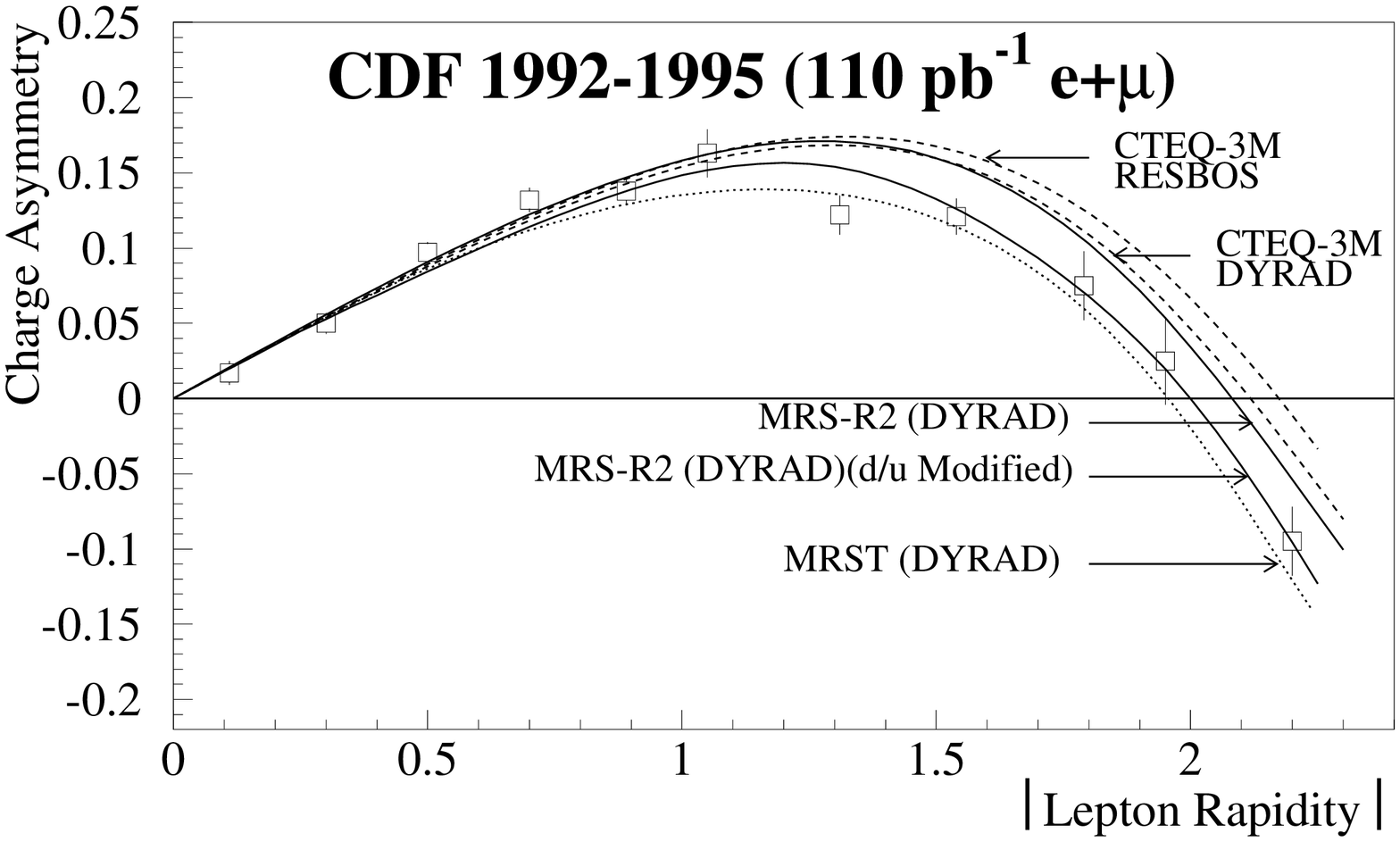,width=12cm}
\caption{CDF lepton charge asymmetry from $W$ decays as a function of 
lepton rapidity. The error bars correspond to statistical and systematic errors
added in quadrature.
\label{wasm}}
\end{figure}

As can be seen in Fig.~\ref{mrs1}, the $u$ quarks in the proton 
carry more momentum on average than the $d$ quarks. (The $\bar{d}$
quarks carry only slightly more momentum than the $\bar{u}$ quarks 
at high $x$.)
$W^+$ and $W^-$ production
in $\bar{p}p$ collisions is primarily due to the annihilation
processes $u(p)\bar{d}(\bar{p})\rightarrow W^+$ and 
$\bar{u}(\bar{p})d(p)\rightarrow W^-$.
On average the $W^+$'s are therefore produced relatively more forward than the 
$W^-$'s.
CDF~\cite{CDFW} measure 
the lepton asymmetry from the $W$ decays, 
$$
A(y_l)=\frac{d\sigma^+/dy_l-d\sigma^-/dy_l}
            {d\sigma^+/dy_l+d\sigma^-/dy_l},
$$
where $d\sigma^\pm/dy_l$ are the cross-sections for $W^\pm$
decay leptons as a function of lepton rapidity relative to 
the proton beam direction. The $W$ decay to leptons 
tends to reduce the asymmetry
at forward $|y_l|$ but in a well-defined manner
(assuming the $W$ decays proceed via a SM {\em V-A} interaction).
In Fig.~\ref{wasm}, the final results from runs Ia and Ib are shown compared
to various PDF's using the DYRAD NLO calculations. The best agreement with the 
data is obtained using the modified MRS(R2) or the MRST PDF (not shown), 
which incorporates this data as input, indicating that the ratio of
$d/u$ is larger than previously thought.

\noindent{\em $\bar{d}/\bar{u}$ at intermediate $x$:}
The structure function measurements constrain the sum of $\bar{d}$ and 
$\bar{u}$ but not the difference.
The classic method to constrain this difference 
is via the Gottfried sum rule where
$$S_G = \int_0^1 F_2^p - F_2^n \frac{dx}{x} = \frac{1}{3} \int_0^1 
(d_v - u_v) dx - \frac{2}{3} \int_0^1 (\bar{d} - \bar{u}) dx.$$
The NMC result for $0.004<x<0.8$ 
extrapolated over all $x$ gives $S_G = 0.235 \pm 0.026$, 
significantly below 1/3.~\cite{michele} 
It should be noted however that the deuterium
data used to derive $F_2^n$ are not corrected for the effects noted above.
This correction is estimated to be $-0.013$ 
i.e. 50\% of the uncertainty.~\cite{kumano}

\begin{figure}
\center
\epsfig{file=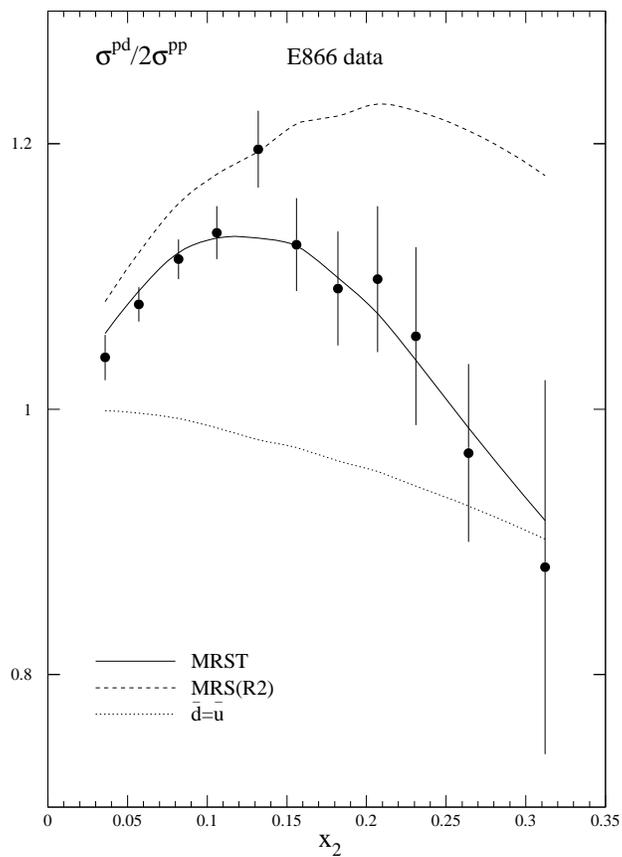,width=9cm}
\caption{E866 ratio of $\sigma^{pp}/2\sigma^{pd}$ versus $x_2$, the momentum
fraction of the target quark, compared to the
NLO calculations discussed in the text.
\label{A866}}
\end{figure}

At this conference, the E866 data were reported,~\cite{E866}  
which provide a significant constraint on the ratio of $\bar{d}/\bar{u}$
via the measurement of Drell-Yan dimuons with mass 
$M_{\mu^+\mu^-} \geq 4.5$~GeV from 800~GeV protons on proton 
and deuterium targets as shown in Fig.~\ref{A866}. Here, 
the ratio of $\sigma^{pd}/2\sigma^{pp}$ is measured as a function
of the reconstructed momentum fraction of the target quark in the
range $0.036 < x_2 < 0.312$. The MRST calculations incorporate a direct
parameterisation of $\bar{d} - \bar{u}$ in order to fit the data. Earlier
PDF's such as MRS(R2) clearly overestimate this asymmetry but a 
$\bar{d}=\bar{u}$ sea is still ruled out. 

\begin{figure}
\center
\epsfig{file=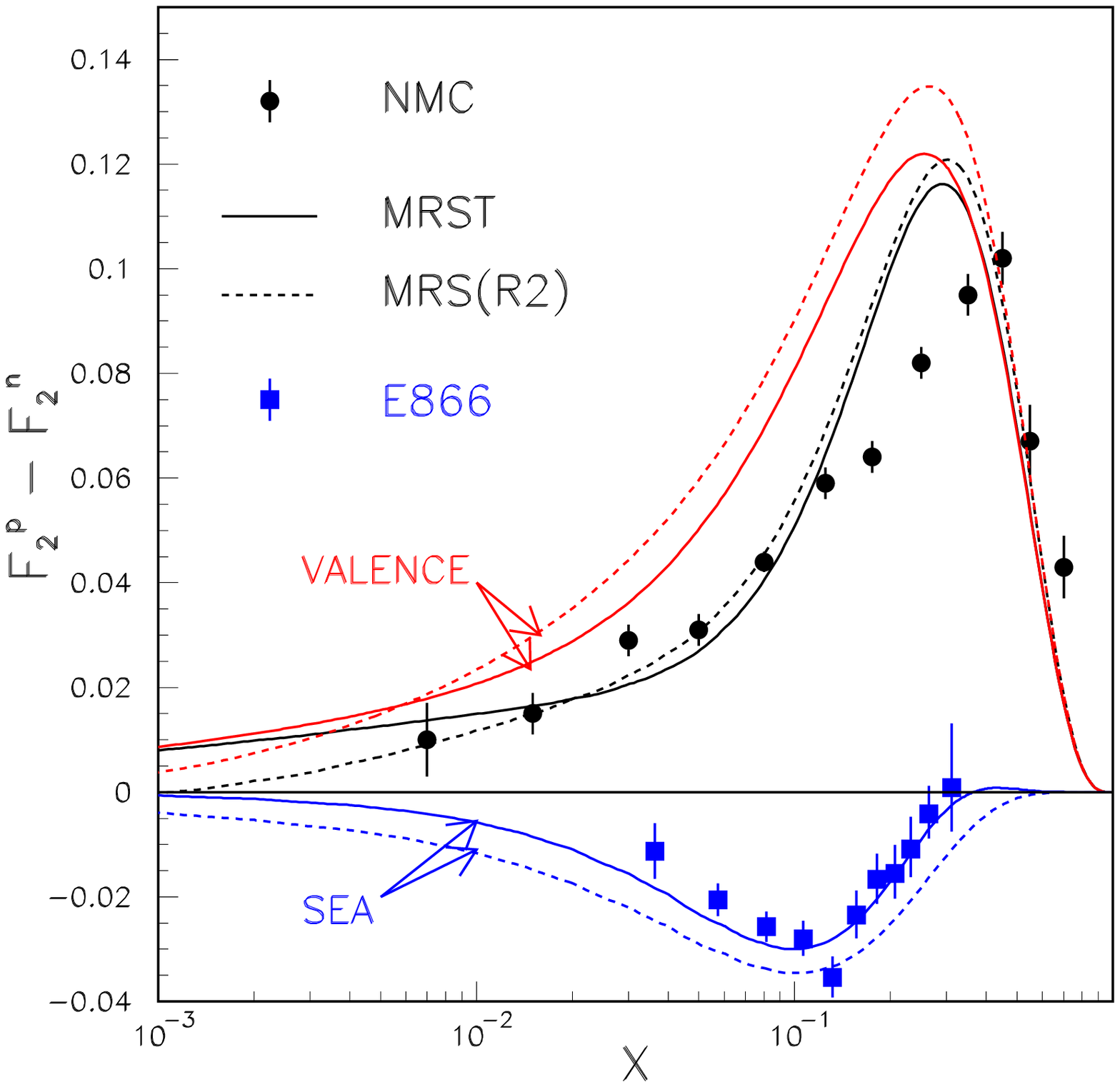,width=10cm}
\caption{NMC $F_2^p - F_2^n$ versus $x$ 
compared to LO predictions discussed in 
the text. 
For each prediction, the top (bottom) curve is the valence (sea) contribution
and the middle curve is the sum.
The E866 results for the sea quark contribution to 
$F_2^p - F_2^n$ are indicated as the negative contribution. 
\label{B866}}
\end{figure}

The E866 data require an overall decrease of the sea
compared to the MRS(R2) parameterisation. This is 
seen in the lower part of 
Fig.~\ref{B866} where  
the E866 data are plotted as the sea contribution
to $F_2^p(x) - F_2^n(x)$,
which contributes negatively to $S_G$, 
evolved to the NMC $Q^2$ values.
The NMC data are plotted in the upper plot, compared to the valence 
as well as the summed contribution
for the MRST and MRS(R2) partons.
Although the valence and sea distributions differ, 
these distributions give similar results for the total $F_2^p(x) - F_2^n(x)$.
However, it is clear that the agreement of the NMC data with either 
parameterisation is poor at intermediate $x$, which would require additional
changes to the valence quark distributions.
The MRST NLO fits determine $S_G = 0.266$, above the NMC measured
value discussed earlier. 
This discrepancy is induced in part by the inclusion of the
E866 data in the fit.

\begin{figure}
\center
\epsfig{file=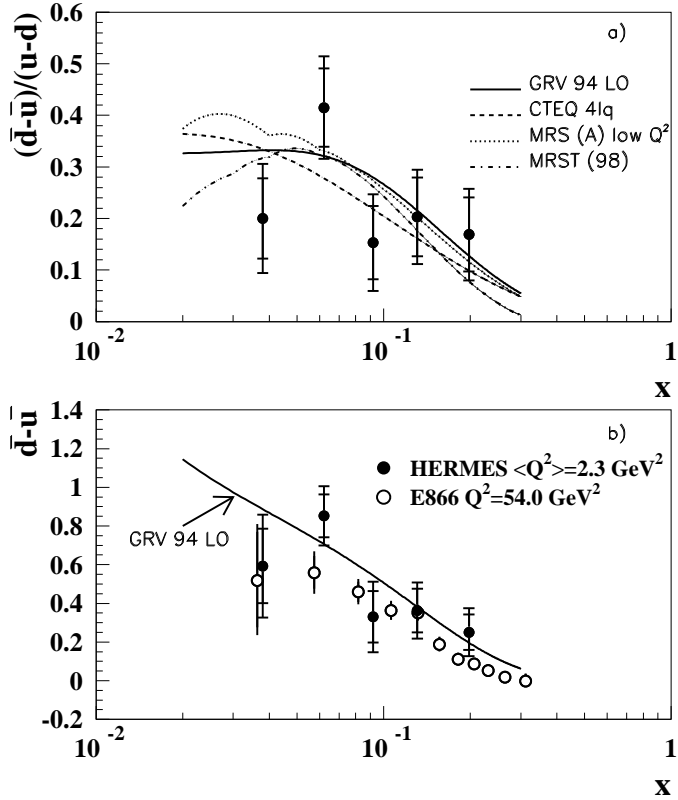,width=10cm}
\caption{a) $(\bar{d} -\bar{u})(u-d)$ extracted 
directly from HERMES data ($\bullet$) compared to 
various parameterisations of the proton. b) HERMES data 
converted to $\bar{d} - \bar{u}$ in order to compare with E866 ($\circ$) data.
\label{HERM}}
\end{figure}

A further constraint is provided via the HERMES 
semi-inclusive measurement of
the ratio of the differences between
charged pion production for (unpolarised) proton and neutron ($^3$He)
targets
$$
r(x,z)= \frac{N^{\pi^-}_p(x,z)-N^{\pi^-}_n(x,z)}{N^{\pi^+}_p(x,z)-
N^{\pi^+}_n(x,z)},
$$
also reported at this conference.~\cite{manuella}
The ratio is observed to scale as a function of the fragmentation scaling 
variable $z$ and the data are used to constrain $(\bar{d}-\bar{u})/(u-d)$
at leading order as a function of $x$, as shown in Fig.~\ref{HERM}a.
Using the PDF's to constrain the valence quarks, the $\bar{d}-\bar{u}$
distribution obtained from the E866 and (lower $Q^2$) HERMES data 
are shown to be in agreement in Fig.~\ref{HERM}b.

Generically, the Pauli exclusion principle, suppresses production of
sea $u\bar{u}$ quarks at large $x$ due to the additional $u_v$ quark.
Hence the excess of $\bar{d}$ over $\bar{u}$ is a direct consequence of
the excess of  $u_v$ over $d_v$. 
The currently favoured dynamical approach is to include non-perturbative 
effects from virtual mesons directly from $|n\pi^+\!>$ Fock states
in a virtual pion model or to derive the sea from the 
valence quarks coupling to Goldstone bosons (e.g. $\pi^+$ from 
$u \rightarrow d\pi^+$) in a chiral model.~\cite{peng}
In conclusion, the latest data indicate that the ratio of $\bar{d}/\bar{u}$
is less than previously thought, a decrease which is correlated 
with the increase of $d/u$.

\noindent{\em $s$ quark contribution:}
The strange contribution to the quark sea 
is determined using opposite-sign 
dimuon events in $\nu Fe$ scattering from CCFR where 
one muon comes from the lepton vertex and the other results from the 
semi-leptonic decay of a charmed hadron.
The ratio of $s$ to $u$ and $d$ quarks integrated over $x$ in an NLO 
analysis was determined to
be $0.477^{+0.063}_{-0.053}$.~\cite{bazarko} 
This large uncertainty requires, for example, 
that NuTeV tag $\nu$ and $\bar{\nu}$ and take the 
difference of these cross-sections in order to reduce the uncertainty on
$sin^2\theta_W$, also reviewed at this conference.~\cite{karlen}
A second potential method to constrain the strange quark contribution
is by comparing $F_{2}^{\nu N}$ and $F_{2}^{l N}$.
The analysis of this data was discussed by 
C. Boros.~\cite{boros}
To leading order the muon (NMC) and neutrino (CCFR) 
structure functions are related by the ``5/18ths rule"
$$F_{2}^{l N} = 
    \frac{5}{18} [ 1 - \frac{3}{5}
      \frac{x s(x) + x \bar{s}(x)}{x q(x) + x \bar{q}(x)}]
         F_{2}^{\nu N}$$
where the ($s\bar{s}$ symmetric) strange sea enters as a correction. 
However, the CCFR data
corrected using the dimuon result to constrain the strange sea 
lies significantly ($\simeq 20\%$) above the NMC data for all values of 
$Q^2$ at $x<0.1$.  
Here nuclear shadowing corrections, estimated independently for $\nu Fe$ 
(CCFR) and $\mu D$ (NMC) data, indicate that part of the discrepancy between 
the two experiments at low $x$ may be accounted for in this way.~\cite{boros}
However, 
there remains a significant discrepancy: the strange quark
sea and the relation between $F_{2}^{\nu N}$ and $F_{2}^{l N}$ data
for $x<0.1$ is therefore an active area of theoretical and
experimental investigation.

\section{Spin Structure Functions}
Major progress has been made in the last two years in the measurement
and QCD analysis of spin structure functions. 
The second generation CERN and SLAC experiments have been 
augmented by the HERMES experiment at DESY and these experiments now 
provide detailed information on the proton, neutron and deuteron
spin structure.
Here, we discuss measurements of the longitudinal asymmetry
$$ A_\| = 
\frac{\sigma^{\downarrow\uparrow}-\sigma^{\uparrow\uparrow}}
     {\sigma^{\downarrow\uparrow}+\sigma^{\uparrow\uparrow}} $$
where $\sigma^{\uparrow\uparrow}$ ($\sigma^{\downarrow\uparrow}$)
is the cross-section when the 
lepton and nucleon spins are parallel (antiparallel) to each other.
This asymmetry takes into account the beam and target polarisations 
as well as the dilution factor, the fraction of polarisable nucleons in
the target.
The asymmetry can then be related 
to the virtual-photon nucleon asymmetries $A_1$ and $A_2$
$$ A_\| = D (A_1 + \eta A_2)$$
where $D$ is the depolarisation factor depending upon $R = \sigma_L/\sigma_T$
and $\eta = \eta(x,Q^2)$ is a kinematic factor.
The spin structure function 
$$g_1 = \frac{F_2}{2x(1+R)} (A_1 + \gamma A_2)$$
where $\gamma$ is a further kinematic factor and $g_1$ is then calculated 
using the known $F_2$ and $R$ values. 
The $A_1$ term dominates and to a reasonable approximation 
the extracted structure function is proportional to the measured asymmetry
$$ g_1 \simeq  \frac{F_2}{2x(1+R)} \cdot A_1 \simeq F_1 \cdot \frac{A_\|}{D}.$$
In the Quark Parton Model
$$ g_1(x, Q^2) = \frac{1}{2} 
\sum_i e_i^2 \cdot \Delta\Sigma(x,Q^2)$$
where $\Delta\Sigma = \sum_i [(q_i^\uparrow - q_i^\downarrow)
+ (\bar{q}_i^\uparrow - \bar{q}_i^\downarrow)]$
is the singlet summed quark and anti-quark distributions 
where $q^\uparrow$ ($q^\downarrow$) 
is the quark distribution with the spins parallel 
(antiparallel) to the nucleon spin.
In the QPM $g_1$ is therefore 
the charge-weighted vector sum of the quark polarisations in the nucleon.
Similarly the non-singlet quark distribution
$\Delta q_{NS} = \sum_i \frac{e_i^2 - <e^2>}{<e^2>}
[(q_i^\uparrow - q_i^\downarrow)
+ (\bar{q}_i^\uparrow - \bar{q}_i^\downarrow)]$,
where $<\!e^2\!> = 2/9$ for 3 light flavours is
defined such that this contribution evolves separately from the 
gluon ($\Delta g$) and singlet ($\Delta\Sigma$) contributions
when QCD corrections are taken into account. 
The DGLAP formalism outlined earlier can therefore be applied
and $\Delta q_{NS}$ constrained by $g_1$ via QCD corrections.

\begin{figure}
\center
\epsfig{file=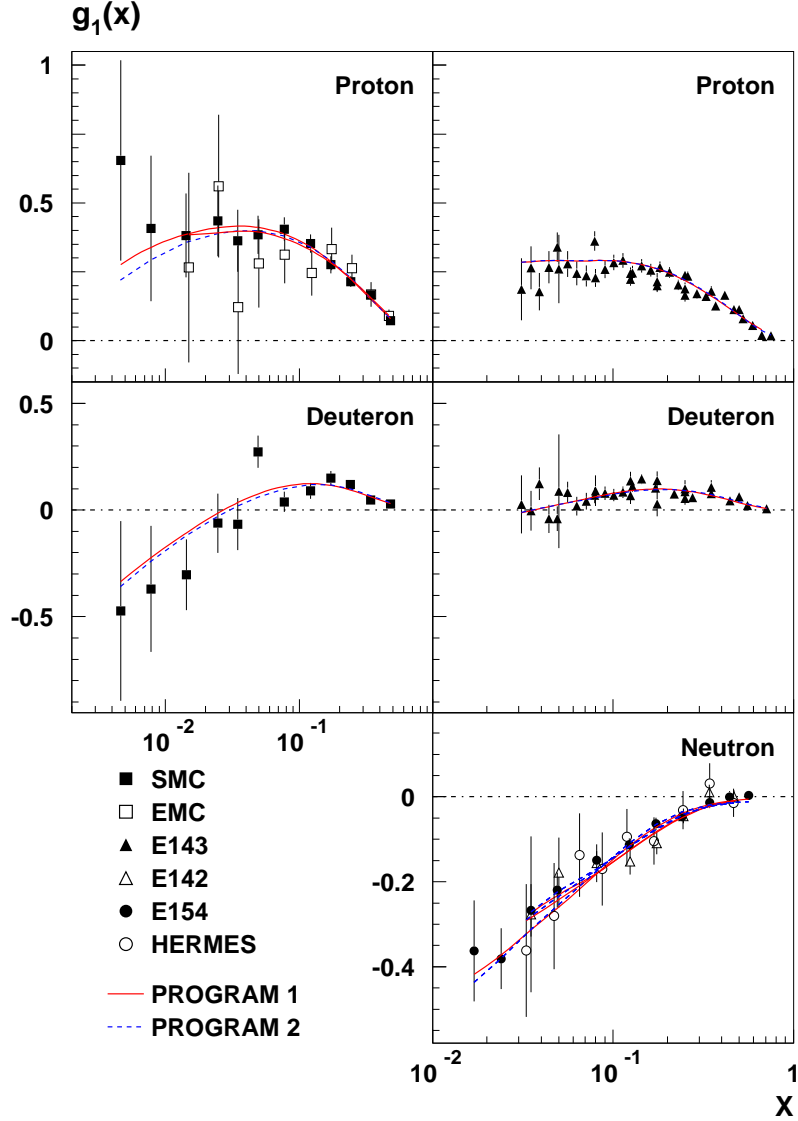,width=12cm}
\caption{Published world data on $g_1(x)$ 
from CERN experiments ($0.003<x<0.6$, $1 <Q^2<100$~GeV$^2$)
in the left column and SLAC
and DESY experiments ($0.014<x<0.7$, $1<Q^2<20$~GeV$^2$)
in the right column with statistical errors only. Results from two
($\overline{\rm{MS}}$ scheme) NLO QCD fitting programs 
discussed in the text are superimposed.
\label{qcdf}}
\end{figure}
The corresponding integrals $\Gamma_1 = \int_0^1 g_1(x, Q^2) dx$
determine the total spin carried by the quarks at the measured $Q^2$.
In the QPM, below charm threshold
$$\Gamma_1^p = 
\frac{1}{2}(\frac{4}{9}\Delta u + \frac{1}{9}\Delta d + \frac{1}{9}\Delta s)$$
and, by isospin symmetry
$$\Gamma_1^n = 
\frac{1}{2}(\frac{1}{9}\Delta u + \frac{4}{9}\Delta d + \frac{1}{9}\Delta s)$$
Beyond the QPM,
the total spin of the nucleon can be written as the sum of the contributions
from its constituents
$$\frac{1}{2} = \frac{1}{2}\Delta\Sigma + \Delta g + L_q + L_g$$
where the $\Delta$'s correspond to the intrinsic spins and the $L$'s correspond
to the angular momentum of the quarks and gluons.
The proton spin puzzle is that only a fraction of 
the total spin is due to quarks. This puzzle remains, 
but first steps have now been taken 
to constrain the gluon contribution via the
scaling violations of $g_1(x)$ and the valence and sea quark contributions
via semi-inclusive measurements.

The world data on $g_1(x)$ for the proton, deuteron and neutron are summarised 
in Fig.~\ref{qcdf} for the CERN, SLAC and DESY experiments.~\cite{abe}
Comparison of the lower and higher $Q^2$ data 
indicates that the scaling violations are relatively small 
over the well-measured range of $1 < Q^2 \sleq 10$~GeV$^2$.
The data are taken using a variety of targets (e.g. 
H($p$), D($d$) and $^3$He($n$) for HERMES;
NH$_3$($p$), ND$_3$($d$), $^6$LiD($d$) and $^3$He($n$) for the SLAC 
experiments;
and, NH$_3$($p$), butanol($p$) and deuterated-butanol($d$) for SMC)
and widely varying experimental techniques.

\begin{figure}[hbt]
\center
\epsfig{file=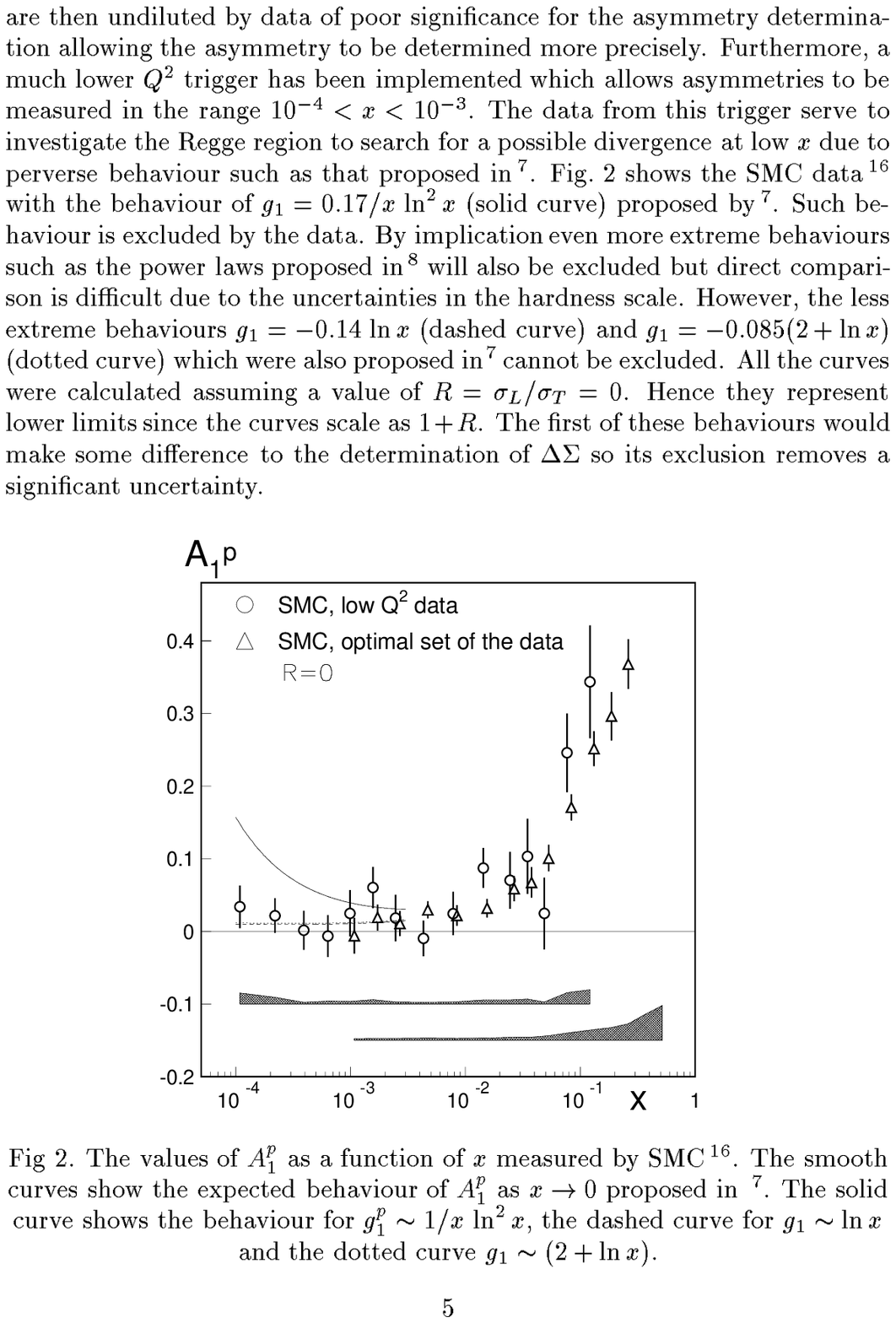,width=9cm,clip=}
\caption{SMC virtual-photon proton asymmetry, $A_1^p$, 
for $<\!Q^2\!> = 0.01$~GeV$^2$ data as a function of $x$ compared to the
behaviours discussed in the text. The systematic errors for the low 
$(x,Q^2)$ and the reference set of data are indicated by the shaded bands.
\label{slox}}
\end{figure}

There is no evidence for a rise of $g_1^p$ at the smallest values of 
$x \sim 10^{-2}$ explored by SMC and E155. Indeed the QCD fit prediction
is that $g_1^p$ becomes negative for $x \sleq 10^{-3}$ due to the
relatively
large positively polarised gluon at higher $x$ 
driving the polarised gluon negative at 
small~$x$.~\cite{altarelli} Clearly this will be of 
interest in the light of the HERA unpolarised results and of significance 
in the extrapolations required for the determination of the sum rules.
SMC presented data on the virtual-photon proton asymmetry, $A_1^p$ at low
$<\!Q^2\!> = 0.01$~GeV$^2$ which extend to very low $x = 10^{-4}$,
as shown in Fig.~\ref{slox}.~\cite{rondio}
The low $x$ data were obtained with a dedicated low-$Q^2$ trigger, 
requiring an observed hadron in each event which 
rejects radiative and other events with low depolarisation factors.
At these very low $Q^2$ values the data indicate that extreme QCD behaviour
of $g_1^p \propto 1/x \ln^2x$ (full line) proposed in~\cite{close}
is ruled out. However the less
extreme QCD behaviours $g_1^p \propto \ln x$ or $2+\ln x$ indicated by the
dotted lines give reasonable descriptions of the data.
Clearly higher $Q^2$ data are desirable in order to test the pQCD models,
but the data do constrain the Regge behaviour. 

The values of $g_1^n$ in Fig.~\ref{qcdf} 
fall monotonically with decreasing $x$.
The precision of this data is now approaching that of the proton data,
which is important in the context of the Bjorken sum rule discussed below.
All data are observed to be in good agreement, where the 
systematic errors (not shown) are typically smaller than the 
statistical errors.
In addition, E155 presented new preliminary
data on $g_1^p$ and $g_1^n$ (not shown) at $Q^2 = 5$~GeV$^2$
which are in good agreement with these published datasets.~\cite{sorrell}
These data have very small statistical errors 
and extend the $x$ range compared to the E143 data.
This improved precision also requires
that nuclear effects due to the assumed 
superposition of D and $^4$He states in the $^6$Li target data are understood, 
but possible uncertainties can be tested by comparison with data from 
different targets and other experiments.
\begin{figure}[hbt]
\center
\hspace*{2cm}\epsfig{file=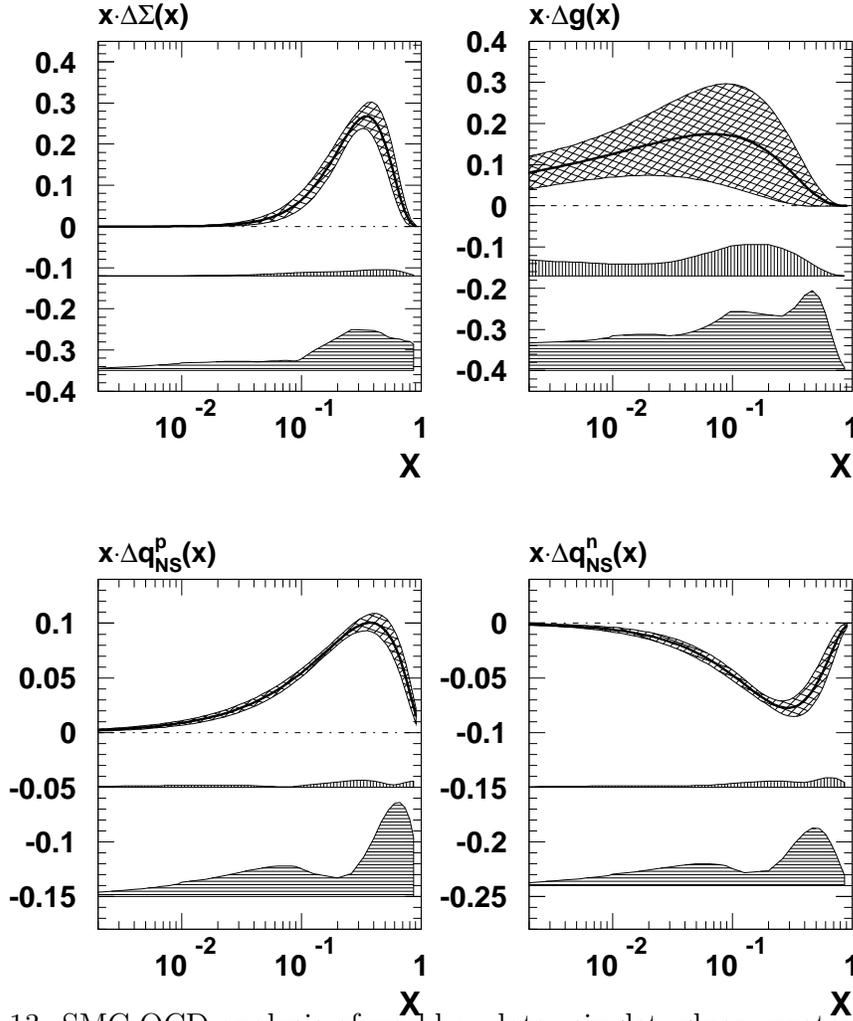,width=14cm}
\caption{SMC QCD analysis of world $g_1$ data: 
singlet, gluon, proton and neutron non-singlet 
polarised parton distributions in AB 
scheme at $Q_{\rm o}^2= 1$~GeV$^2$.
The fits with the statistical uncertainty are indicated by the cross hatched
upper bands. The (small) experimental systematic uncertainties are indicated 
by the central bands and the (larger) theoretical uncertainties by the
lower bands.  
\label{SMCf}}
\end{figure}

SMC have performed a NLO QCD fit~\cite{abe} 
to extract the singlet ($x\cdot\Delta\Sigma$)
and gluon ($x\cdot\Delta g$)
as well as the proton ($x\cdot\Delta q_{NS}^p$) 
and neutron ($x\cdot\Delta q_{NS}^n$) 
non-singlet polarised parton distributions shown in Fig.~\ref{SMCf}.
Here, the deuteron spin 
structure function $g_1^d$ is assumed to be related to the 
proton and neutron structure functions by
$$ g_1^p + g_1^n = \frac{2g_1^d}{1-\frac{3}{2}\omega_D}$$
where $\omega_D = 0.05 \pm 0.01$ is the D-wave state probability in the 
deuteron. 

The results of the fits from two NLO QCD programs are
shown by the full and dashed lines in Fig.~\ref{qcdf} at the 
measured $Q^2$ of each of the datasets.
The comparison indicates that a 
good fit is obtained to the world data with $\chi^2/DoF = 127.4/(133-8)$,
considering statistical errors only.
The parton densities are parameterised and determined in analogy to the 
unpolarised case via the NLO DGLAP splitting kernels. 
Results are 
quoted in the Adler-Bardeen (AB) scheme, a modified version of 
the more conventional $\overline{\rm{MS}}$ scheme, defined such that
$\Delta\Sigma_{\rm{AB}}$ is independent of $Q^2$.
These renormalisation/factorisation schemes are
related via

\begin{eqnarray}
a_0(Q^2) & = & \Delta\Sigma_{\rm{\overline{MS}}}(Q^2) \nonumber
\\
           & = & \Delta\Sigma_{\rm{AB}} - 
n_f\frac{\alpha_s(Q^2)}{2\pi}\Delta g_{\rm{AB}}(Q^2) \nonumber
\end{eqnarray}
where $a_0(Q^2)$ is the singlet axial-current matrix element and
the $\Delta$'s correspond to the parton densities
integrated over $x$. This scheme dependence is large and sufficient
to create a negative $\Delta\Sigma(x)$ at small $x$ and a smaller 
$\Delta g(x)$ in $\overline{\rm{MS}}$ scheme.~\cite{abe}
However, the physical structure functions and their integrals, such
as $a_0(Q^2)$ are
unaffected by this definition. 
In Fig.~\ref{SMCf}, the singlet contribution is seen to be well constrained
by $g_1^p$. 
This remains true when account is taken of the  theoretical uncertainties
which include the
variation of the renormalisation and factorisation scale by factors of two,
varying $\alpha_s = 0.118\pm0.003$ within the given limits and changes to
the starting scale and the functional forms of the parameters.
Similarly the non-singlet contributions 
are reasonably well
constrained and the rise of $\Delta q_{NS}^p$ is mirrored by the fall of
$\Delta q_{NS}^n$. 
However, $\Delta g$ is rather poorly constrained especially
when taking into account the theoretical uncertainties.
Integrating over $x$ at $Q_{\rm o}^2 = 1~{\rm GeV}^2$
$$
\tiny
\Delta g = 0.99_{-0.31}^{+1.17} (stat.)
_{-0.22}^{+0.42} (sys.) _{-0.45}^{+1.43} (theory)$$
in AB scheme compared to $\Delta g = 0.25_{-0.22}^{+0.29} (stat.)$ in 
$\overline{\rm{MS}}$ scheme. 
Consistent results are obtained in either scheme for the singlet contribution
expressed in terms of the axial-current matrix element
$$a_0(Q_{\rm o}^2 = 1~{\rm GeV}^2) = 0.23 \pm 0.07 (stat.) \pm 0.19 (sys.)$$
which compares to the QPM expectation $\simeq 0.58$ and
corresponds to about one third of the nucleon spin being carried by quarks.

In the latest analysis by J. Ellis and M. Karliner,~\cite{ellis} 
the world average value
of the singlet matrix element in $\overline{\rm{MS}}$ scheme is 
given by
$$ \Delta\Sigma = 0.27 \pm 0.05.$$
This approach utilises the $\mathcal{O}$$(\alpha_s^3)$ calculations and 
$\mathcal{O}$$(\alpha_s^4)$ estimates discussed below
in relation to the Bjorken sum rule.
Here, the consistency of the data taken at different $Q^2$ values improves 
as successive higher-order QCD corrections are taken into account.

\noindent{\em Bjorken Sum Rule:}
The Bjorken sum rule is a fundamental prediction of QCD 
determined by the difference of the spins carried
by the $u$ and the $d$ quarks
$$ \Gamma_1^p - \Gamma_1^n = \int_0^1 (g_1^p - g_1^n) dx 
= \frac{1}{6}|\frac{g_A}{g_V}| 
\cdot C_1^{NS}(Q^2)$$
where $g_A$ and $g_V$ are the axial-vector and vector weak coupling constants
of neutron $\beta$-decay and $C_1^{NS}(Q^2)$ is the 
non-singlet perturbative QCD correction which
has been calculated up to $\mathcal{O}$$(\alpha_s^3)$ 
and estimated up to $\mathcal{O}$$(\alpha_s^4)$ using the same approach
as for the GLS sum rule.~\cite{larin,kataev}
Here, higher twist 
contributions to the difference of the structure functions are assumed to 
be negligibly small.

In the SMC NLO fits discussed above, the sum rule is imposed as an input with
$|\frac{g_A}{g_V}|$ fixed rather precisely from $\beta$-decay experiments.
However, the NLO fits provide two methods to check 
the Bjorken sum rule. First, the input can be relaxed and
$|\frac{g_A}{g_V}|$ fitted as an additional parameter in the global fit
yielding
$$\tiny \Gamma_1^p - \Gamma_1^n = 0.174 \pm 0.005 (stat.)
_{-0.009}^{+0.011} (sys.)  
_{-0.006}^{+0.021} (theory)$$
at $Q^2 = 5~{\rm GeV}^2$,
which is in agreement with the theoretically expected 
value of $0.181\pm 0.003$.
Second, the non-singlet contributions $g_1^p - g_1^n$ can be
evolved to a common $Q^2$ in order to determine the sum rule more directly
and with a minimum number of parameters. This yields a consistent value
of $0.181_{-0.021}^{+0.026} (total)$ when evaluated at $\mathcal{O}$$(\alpha_s^2)$.
This second method is potentially more precise, but awaits the high 
statistics data on $g_1^p$ from E155 combined with the existing data on
$g_1^n$ from E154.

\begin{figure}
\center
\epsfig{file=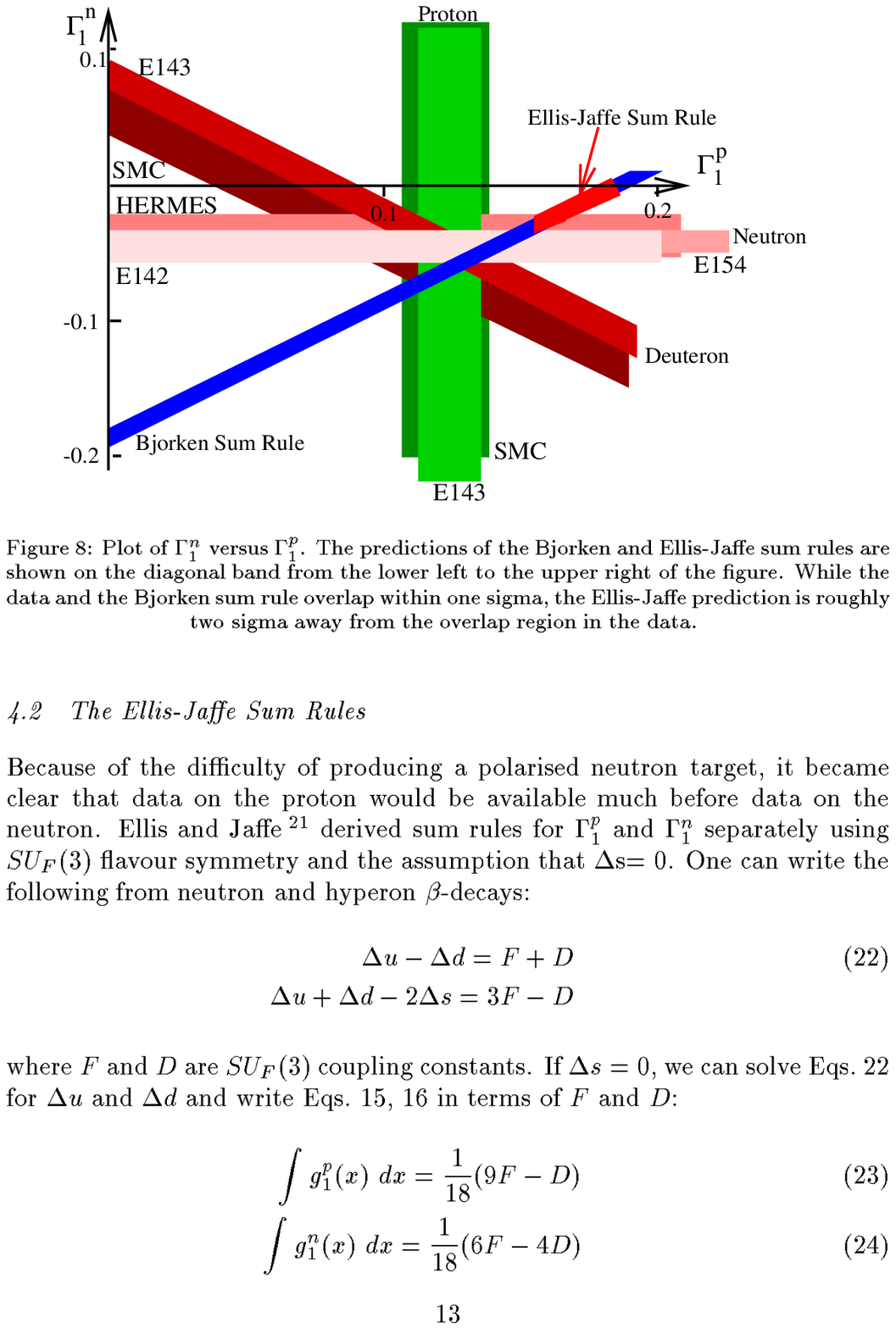,width=15cm,clip=}
\caption{Spin sum rules $\Gamma_1^p$ versus $\Gamma_1^n$. Proton 
(E143 and SMC) 
and neutron (E142, E154 and HERMES) data are indicated by the 
vertical and horizontal bands, respectively.
Deuteron (E143 and SMC) data are 
indicated by the falling diagonal bands. 
The theoretical expectation from the Bjorken sum rule 
($=\Gamma_1^p - \Gamma_1^n$) is indicated by the rising diagonal band
while the Ellis-Jaffe sum rule is the area on this band indicated by
the arrow.
\label{sum1}}
\end{figure}
\begin{figure}
\center
\epsfig{file=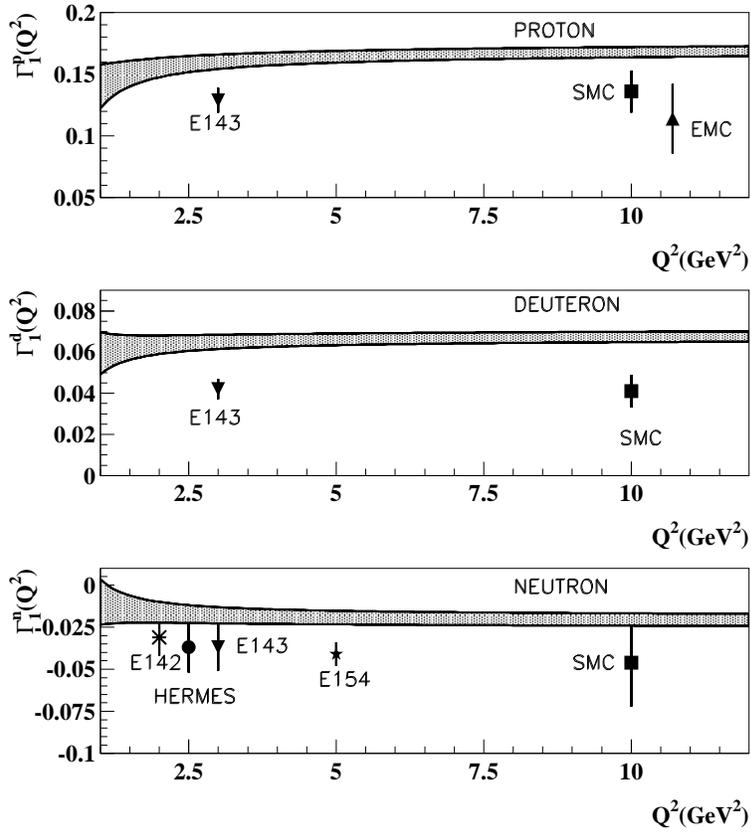,width=10cm}
\caption{Spin integrals, $\Gamma_1$ proton, deuteron and neutron data 
versus $Q^2$ compared to the Ellis-Jaffe sum rule ($\Delta s = 0$ and SU(3)$_f$
symmetry) expectation indicated by the shaded band.
\label{ejsr}}
\end{figure}
\noindent{\em Ellis-Jaffe Sum Rule:}
Assuming 
that strange quarks do not contribute to the nucleon spin
and SU(3)$_f$ symmetry, 
Ellis and Jaffe derived independent
sum rules for the proton and neutron
$$\Gamma_1^p = \frac{1}{2}(\frac{4}{9}\Delta u + \frac{1}{9}\Delta d)
\simeq 0.17$$
$$\Gamma_1^n = \frac{1}{2}(\frac{1}{9}\Delta u + \frac{4}{9}\Delta d)
\simeq -0.02$$
in the high-$Q^2$ limit and modified by singlet and non-singlet QCD 
corrections. 
In Fig.~\ref{sum1} the world data on the Bjorken and Ellis-Jaffe sum 
rules are depicted graphically. As noted in relation to the SMC analysis,
the data are in agreement and consistent with the Bjorken sum rule with a 
precision of around 10\%. 
The origin of the spin puzzle was the EMC measurement of 
$\Gamma_1^p$. In Fig.~\ref{ejsr},
the world data on $\Gamma_1^p$, $\Gamma_1^d$ and $\Gamma_1^n$
all indicate that the Ellis-Jaffe sum rule is broken at the 2-3$\sigma$ level.
The strange sea quarks and/or the gluon 
therefore carry a significant fraction of the spin.
It is currently impossible to distinguish 
an SU(3)$_f$ symmetric sea or a $\Delta s = 0$ (large gluon) 
solution in the NLO QCD fits.~\cite{lampe}
A natural 
assumption would be that SU(3)$_f$ symmetry is violated at the same level
as in the unpolarised structure functions and the rest of the spin
can be attributed to a large gluon polarisation, but this requires
further experimental input. 

\begin{figure}
\center
\epsfig{file=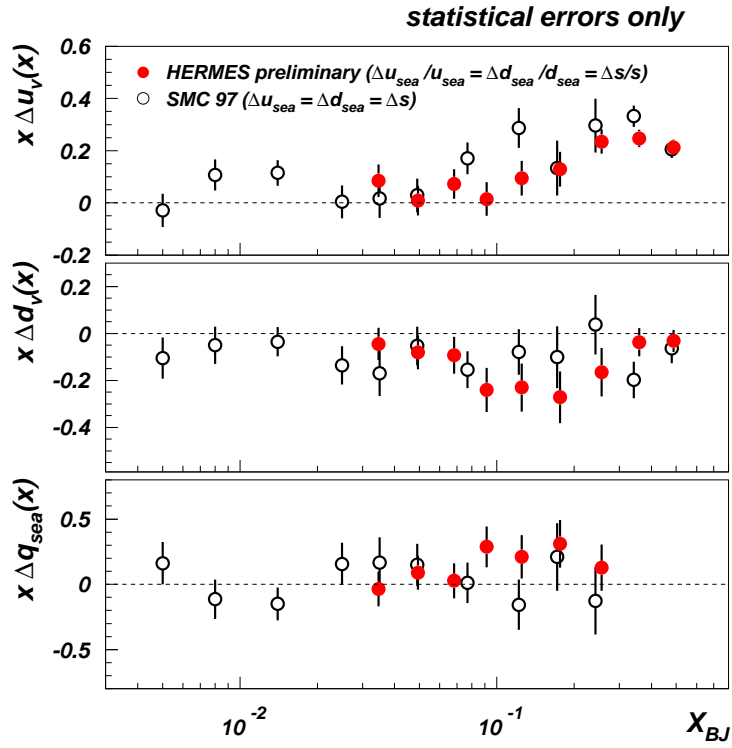,clip=,width=10cm}
\caption{Data from SMC and HERMES for 
the valence ($\Delta u_v$ and $\Delta d_v$) and sea ($\Delta \bar{q}$)
spin densities versus $x$ 
extracted from semi-inclusive asymmetry measurements.
The dominant statistical errors only are shown.
\label{sias}}
\end{figure}

\noindent{\em Semi-inclusive Asymmetries:}
SMC~\cite{SMCsemi} and HERMES~\cite{miller} 
have recently produced data tagging the charge of
final state hadrons. The asymmetry
$A_\|^h(x,z)$ is defined in analogy to the inclusive case 
where the distribution of charged hadrons 
with momentum fraction $z$ 
is statistically correlated with the struck quark
flavour in a flavour tagging analysis.
In Fig.~\ref{sias} the extracted spin contributions $\Delta u_v$,
$\Delta d_v$ and $\Delta \bar{q}$ (introduced for $x < 0.3$) from
SMC and HERMES are observed to be in agreement.
Slightly different assumptions are made with respect to the strange
quark sea where SMC assume an SU(3)$_f$ symmetric sea and HERMES
assume this symmetry is violated at the same level
as in the unpolarised structure functions. However these effects are negligible
for the predominantly pion final states in this
LO QCD analysis. Both experiments observe 
$\Delta u_v$ is positive and $\Delta d_v$ is negative.
The SMC results integrated over $x$ are 
\begin{eqnarray}
\Delta u_v & = & +0.77\pm0.10(stat.)\pm0.08(sys.) \nonumber
\\
\Delta d_v & = & -0.52\pm0.14(stat.)\pm0.09(sys.) \nonumber
\\
\Delta \bar{q} & = & +0.01\pm0.04(stat.)\pm0.03(sys.) \nonumber
\end{eqnarray}
in agreement with the expectations from the NLO QCD fits.
The polarised sea is compatible with zero although there are indications from 
the HERMES data that this contributes positively.

\noindent{\em Outlook:}
The analysis of polarised structure function data enables  
first measurements of the scaling violations to be performed. 
These provide a test of QCD
and indicate that the gluon polarisation is positive. Similarly, the
semi-inclusive measurements give first constraints on the sea.
The structure function data from E155 and semi-inclusive data from HERMES
will soon provide further input.
These data are, however, insufficient to 
determine the various spin parton distributions
within the nucleon. By analogy with Table~\ref{tab:mrst}, 
we can therefore consider
a future programme of spin structure measurements which will enable the 
partonic spin structure to be unravelled, as shown in Table~\ref{tab:thom}. 
In the next few years HERMES, 
COMPASS and the RHIC experiments will focus on the determination of the gluon. 
It is also possible that the
polarised proton technology developed at RHIC could be utilised at HERA
in order to explore polarised structure functions at low $(x,Q^2)$.
With a high luminosity HERA machine the data would also extend 
to high $Q^2$ where neutral current and charged current events
would provide detailed information on this spin structure.

\begin{table*}[htb]
\caption{Processes studied in global spin parton distribution fits.
The first group of datasets correspond to current experiments.
The second group correspond to near-future (year 2000) experiments.
The third group correspond to potential year 2005 experiments
at HERA. (Courtesy of T. Gehrmann.)
\label{tab:thom}}
\vspace{0.2cm}
\begin{center}
\begin{tabular}{|l|l|l|}    \hline
{\bf Process/}     &     {\bf Leading order}   & 
{\bf Parton behaviour probed}\\
{\bf Experiment}   &  {\bf subprocess}         &                           \\
\hline
&\hfill \raisebox{-0.5ex}[0.5ex][0.5ex]{}&                      \\ 
{\bf DIS} $\mbox{\boldmath $(\ell N \rightarrow \ell X)$}$ &  $\gamma^*q 
\rightarrow q$ \hfill \hspace*{4pt}& 
 Two structure
functions $\rightarrow$ \\ 
$g^{\ell p}_1,g^{\ell d}_1,g^{\ell n}_1$
& \hfill \hspace*{4pt}& $\sum_q e_q^2 (\Delta q+\Delta \bar{q})$ 
\\ 
(SLAC, EMC/SMC,& \hfill \hspace*{4pt}& 
 $\Delta A_3 = \Delta u+\Delta \bar{u}-\Delta d-\Delta \bar{d}$ \\   
HERMES)& \hfill \hspace*{4pt}& 
\hspace*{1cm}  \\ 
      &\hfill  & \hspace*{1cm}    \\
$\mbox{\boldmath $\ell p, \ell n \rightarrow \ell \pi X$}$    &  
$\gamma^* q \rightarrow q$         
with   &   $\Delta u_v$, $\Delta d_v$, $\Delta \bar{q}$\\
(SMC,HERMES)   & $q = u, d, \bar{u}, \bar{d}$ &     \\[2mm] \hline   
     &     &    \\    
$\mbox{\boldmath $\ell N \rightarrow c \bar{c} X$}$ & 
  $\gamma  g\rightarrow  c\bar c$    & $\Delta g$ \\ 
 (COMPASS, HERMES)            &     &     $(x\approx 0.15, 0.3)$
\\[2mm]
     &     &    \\    
$\mbox{\boldmath $\ell N \rightarrow h^+h^- X$}$ & 
  $\gamma  g\rightarrow  q\bar q$    & $\Delta g$ \\ 
 (COMPASS)            &  $\gamma  q\rightarrow  q g$   &  $(x\approx 0.1) $ 
\\[2mm]
     &     &    \\    
$\mbox{\boldmath $pp \rightarrow (\gamma^{*},W^{\pm},Z^{0}) X$}$ & 
  $q \bar q \rightarrow \gamma^{*},W^{\pm},Z^{0}$    
& $\Delta u$, $\Delta \bar{u}$, $\Delta d$, $\Delta \bar{d}$ \\ 
 (RHIC)            &     & ($x\gapproxeq 0.06$)     
\\[2mm]
     &     &    \\    
$\mbox{\boldmath $pp \rightarrow \mbox{{\bf jets }} X$}$ & 
  $q \bar q, qq, qg, gg \rightarrow 2j$    
& $\Delta g$ (?) \\ 
 (RHIC)            &     &     (x\gapproxeq 0.03)  
\\[2mm]
     &     &    \\    
$\mbox{\boldmath $pp \rightarrow \gamma X$}$ & 
  $qg \rightarrow q\gamma$    
& $\Delta g$ \\ 
 (RHIC)            &  $q\bar q \rightarrow g\gamma$    & (x\gapproxeq 0.03) 
\\[2mm]\hline 
     &     &    \\    
{\bf DIS} $\mbox{\boldmath $(e^{\pm} p \rightarrow \nu X)$}$ & $W^*q \rightarrow 
q^{\prime}$  \hfill\hspace*{4pt}& Two structure
functions $\rightarrow$ 
\\ 
$g^{\pm}_1,g^{\pm}_5$    &\hfill 
   \hspace*{4pt}& $\Delta u - \Delta \bar{d} - \Delta \bar{s}$
 \\
(HERA)     &\hfill \hspace*{4pt}& $\Delta d + \Delta s - \Delta \bar{u}$
  \\ [2mm]
    &     &    \\   
{\bf DIS (small $x$)}    &   $\gamma^* q \rightarrow q$   &   $\alpha_{q,g}$ 
    \\ 
$g^{ep}_1$ (HERA)   &                  & $(\Delta \bar{q} \sim x^{\alpha_q}, 
 \Delta g \sim  x^{\alpha_g})$    \\[2mm]  
     &     &    \\    
$\mbox{\boldmath $\ell N \rightarrow \ell \mbox{{\bf  ~jets~}} X$}$ & 
  $\gamma^{\star}  g\rightarrow  q\bar q$    & $\Delta g$ \\ 
 (HERA)            &  $\gamma^{\star}  q\rightarrow  q g$   &  $(x\gapproxeq 0.0015) $ 
\\[2mm]
     &     &    \\ 
$\mbox{\boldmath $pp \rightarrow \ell^+ \ell^- X$}$ & 
  $q \bar q \rightarrow \gamma^{\star}$    
& $\Delta \bar{q}$ \\ 
 (HERA-$\vec{{\rm N}}$)            &     & ($x\gapproxeq 0.15$)     
\\[-6pt]
     &     &    \\    
   \hline
\end{tabular}
\end{center}
\end{table*}

\section{HERA Results} 
\begin{figure}[hbt]
\center
\epsfig{file=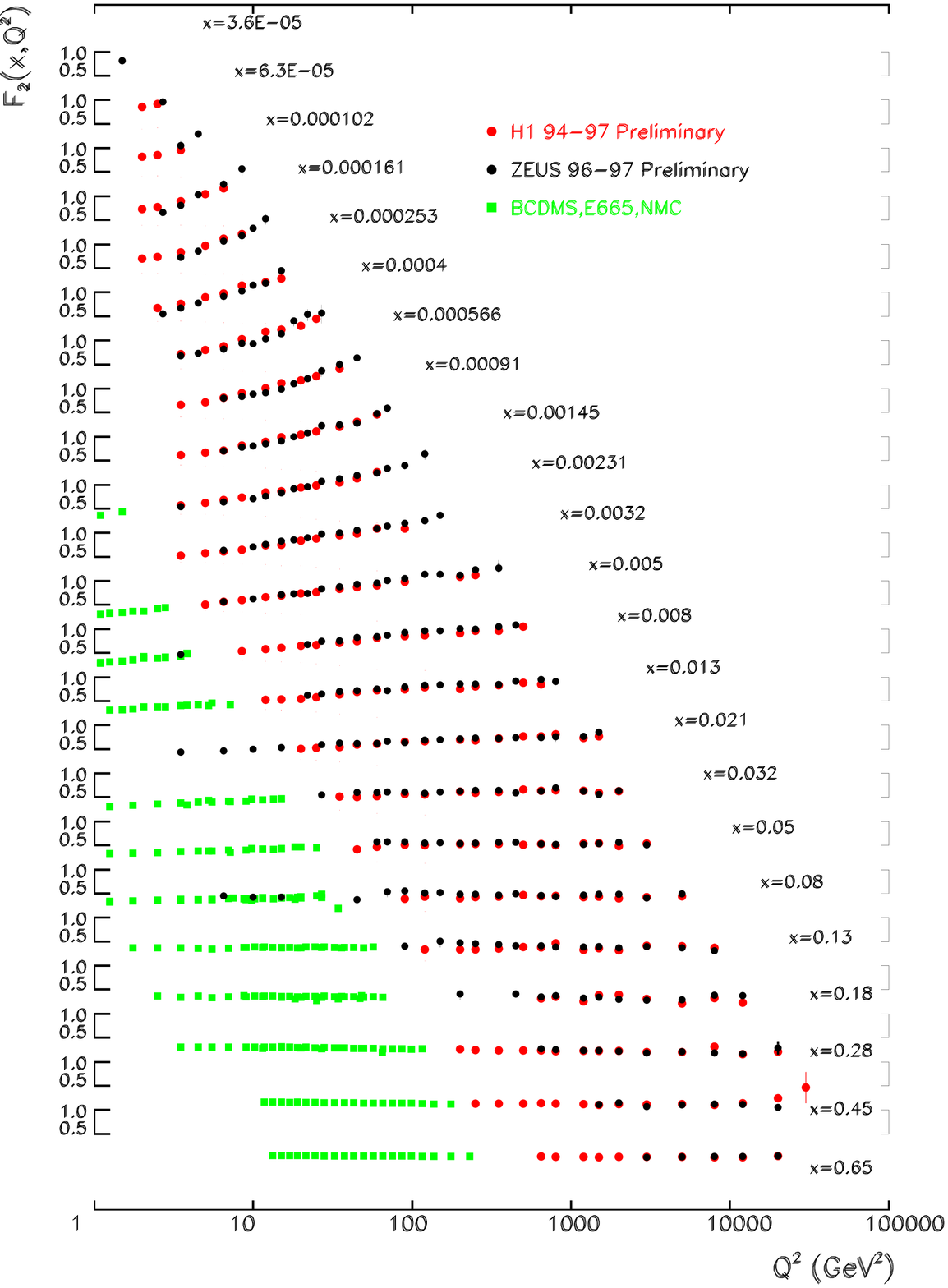,width=12cm}
\caption{Scaling violations of $F_2$ versus $Q^2$ for various $x$ ranges
from the latest H1 and ZEUS preliminary data compared to fixed-target 
experiments. (The quoted $x$ values correspond to the ZEUS measurements
or the nearest $x$ values from other experiments).
\label{f2vq}}
\end{figure}
The HERA $ep$ collider has improved its performance in successive years,
providing large $e^+p$ datasets from 1994-97 running. 
During this period 27.5~GeV positrons collided with 820~GeV protons 
resulting in data samples of 37pb$^{-1}$ and 46pb$^{-1}$ for 
H1 and ZEUS, respectively.
As a measure of the progress which has been made in the study of the
scaling violations of $F_2$, Fig.~\ref{f2vq} illustrates how the latest
HERA data, presented for the first time at ICHEP98, extend the reach in 
$Q^2$ beyond 10,000 GeV$^2$ with $x$ extending up to 
0.65. At high $Q^2$ values, the effects of $Z^0$ exchange are significant
and the value of $F_2^{em}$ is quoted where
$$F_2 = F_2^{em} + \frac{Q^2}{Q^2+M_Z^2}F_2^{int} +
\left(\frac{Q^2}{Q^2+M_Z^2}\right)^2 F_2^{wk}$$
where $F_2^{em}$, $F_2^{int}$ and $F_2^{wk}$ are the contributions due
to photon exchange, $\gamma Z$ interference and $Z$ exchange, respectively.
The data also extend to very low $x$ below $10^{-5}$: here 
it should be noted that the precision of the low-$x\sleq0.03$ 
data is now comparable to that of the fixed-target data at higher $x$.
There is a region of overlap where the HERA and fixed-target 
experiments can be compared.
In particular, the analysis of the ZEUS results at low $y$ may
resolve the CCFR-NMC discrepancy discussed earlier in the context of
the ``5/18ths rule". However, this comparison 
(not shown) is currently inconclusive, 
with the ZEUS data lying between the NMC and CCFR data.

\subsection{Low-$Q^2$ Results}
\begin{figure}[hbt]
\center
\epsfig{file=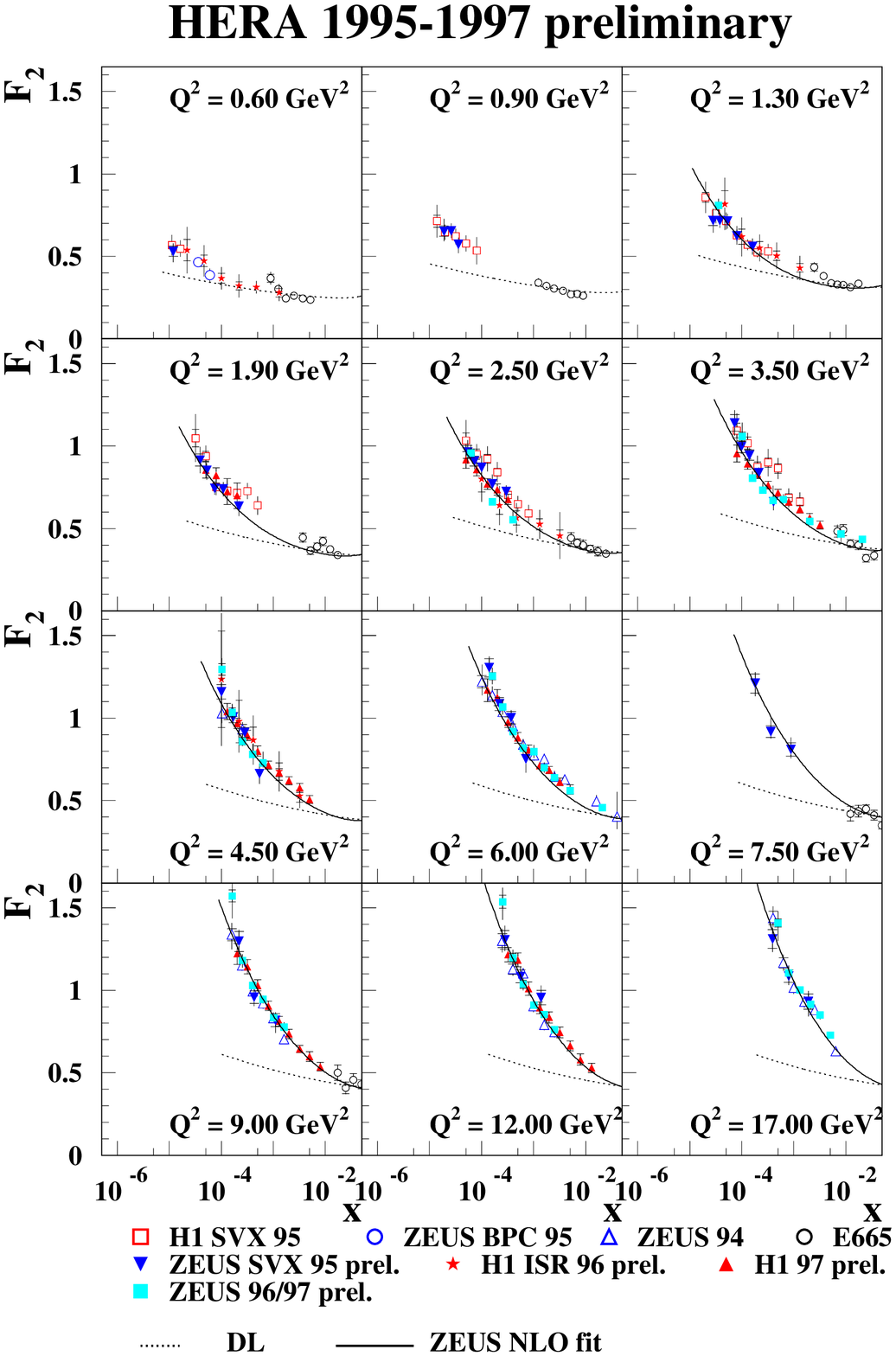,
bbllx=34pt,bblly=27,bburx=529,bbury=790,
width=12cm}
\caption{HERA $F_2$ data for various $Q^2$ intervals as a function of 
$x$ exploring the transition region. The data are compared to the 
Donnachie-Landshoff Regge model (lower dotted line) and the ZEUS NLO 
QCD fit (full line) discussed in the text.   
\label{xn01}}
\end{figure}

\noindent{\it Transition Region:} 
The rise of $F_2$ with decreasing $x$ or, equivalently,
the rise of $\sigma^{tot}_{\gamma^{\star} p}$ with increasing $W$  
has stimulated significant experimental and theoretical developments
in the understanding of QCD at high energies. One challenge is to
explore how and where the transition occurs from soft to hard physics
and interpret low-$Q^2$ data. 
Measurements have been performed using dedicated low-angle taggers 
(e.g. the ZEUS BPC) and 
shifted vertex (SVX) data~\cite{arnulf} as well as 
QED Initial State Radiation (ISR) data~\cite{max} 
in order to map out this region. In Fig.~\ref{xn01}, a compilation of the 
latest measurements available from HERA and E665 indicates that the different
experiments and techniques agree with a precision of around 5\% from the 
most recent data. The significant rise of $F_2$ is apparent for 
$Q^2 \gsim 1$~GeV$^2$, a behaviour which is described by the ZEUS NLO (pQCD)
fit. 
This behaviour is not reproduced in the DL (Donnachie-Landshoff Regge) fit.
The ZEUS collaboration has performed both types of fits to the $F_2$ data 
exploring this transition region.~\cite{arnulf} 
In order to determine the relationship between low-$Q^2$ 
ZEUS BPC data measured in the range 
$0.1<Q^2<0.65$~GeV$^2$ and
$Q^2=0$~GeV$^2$ data, 
a Generalised Vector Meson Dominance (GVMD) 
approach can be taken.
GVMD relates the virtual-photon cross-section
to the real cross-section via
$$\sigma^{tot}_{\gamma^{\star} p} = \sigma^{tot}_{\gamma p}
\cdot {M^2_0 \over Q^2 + M_0^2}$$
for fixed $W$ and $\sigma_L=0$.
A good description of the data is found with
$M^2_0 = 0.53 \pm 0.04 \pm 0.10$~GeV$^2$.
Regge theory then determines the 
$W$ dependence of the data, combined with lower energy photoproduction
experiments, as
$$\sigma_{\rm tot}^{\gamma p}(W^2) = \Areg (W^2)^{\alphareg-1}+ 
\Apom(W^2)^{\alphapom-1}$$
where the Reggeon intercept $\alphareg$ 
is fixed to 0.5 from hadroproduction data
and lower energy photoproduction 
data also constrain $\Apom$, $\Areg$ and $\alphapom$.
From the BPC data alone, the pomeron intercept value is 
$$\alphapom^{BPC} = 1.141 \pm 0.020(stat.) \pm 0.044(sys.)$$
to be compared with the Donnachie-Landshoff value $\alphapom = 1.08$.
In this $Q^2$ range, the rise of the cross-section is therefore relatively
modest. 
Combining the GVMD and Regge approaches, the resulting ZEUSREGGE fit is used
to parameterise the $Q^2$ and $W$ dependence of the low $Q^2<1$~GeV$^2$ data.
The ZEUS NLO QCD fit to the ZEUS 94 and ZEUS SVX data, incorporating NMC and BCDMS 
data, (discussed below) is used to determine the 
behaviour and the uncertainties on the gluon and singlet quark densities at 
low~$x$ for $Q^2 > 1$~GeV$^2$.
\begin{figure}
\center
\epsfig{file=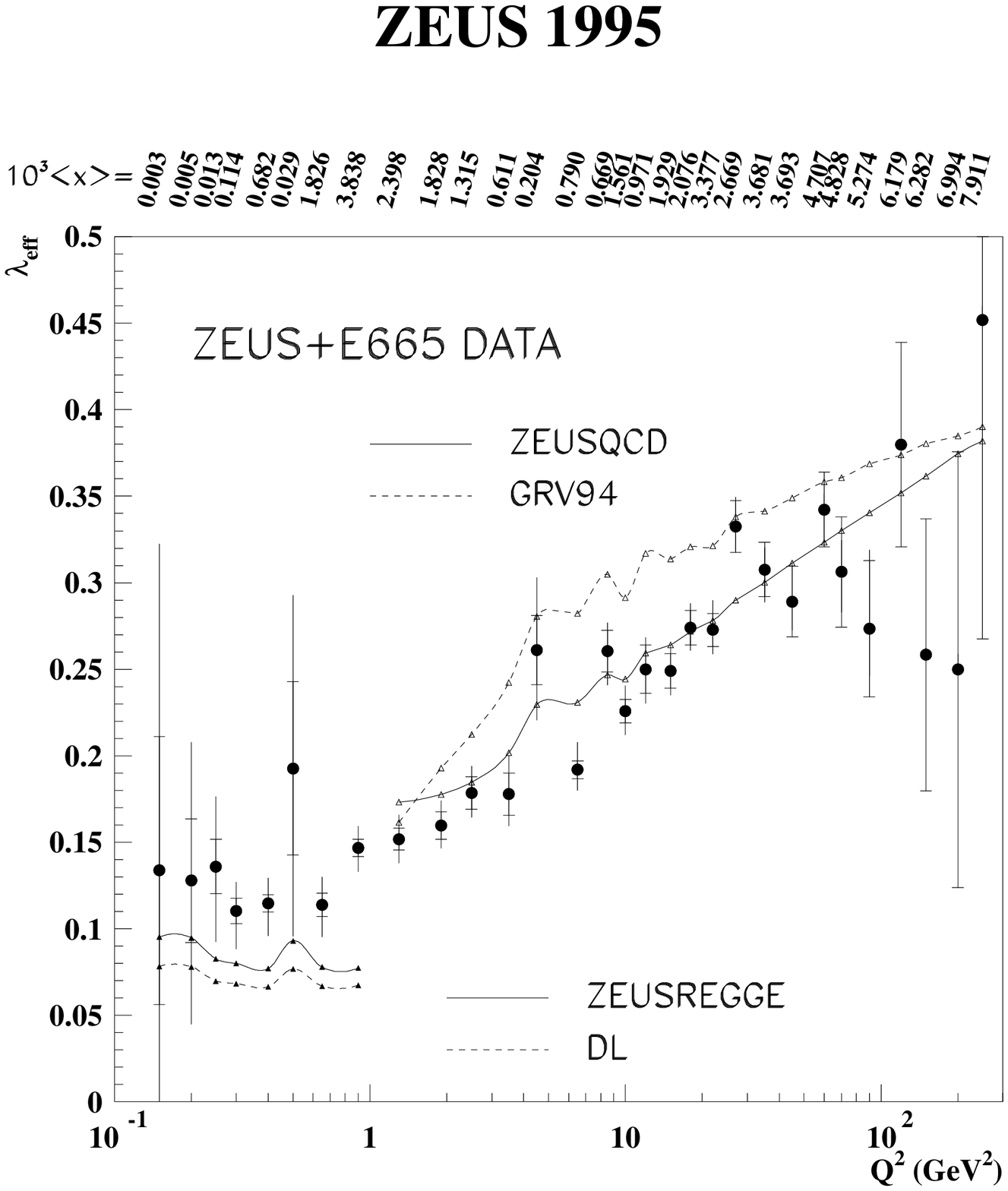,width=7.5cm}
\caption{$\lambda_{\rm eff}$ versus $Q^2$ and $<\! x\!>$
from fits to ZEUS and E665 data
of the form $F_2 = c\cdot x^{-\lambda_{\rm eff}}\,|_{Q^2}$.
The data are compared to the QCD and Regge fits discussed in the text.
\label{xn02}}
\end{figure}
\begin{figure}
\center
\epsfig{file=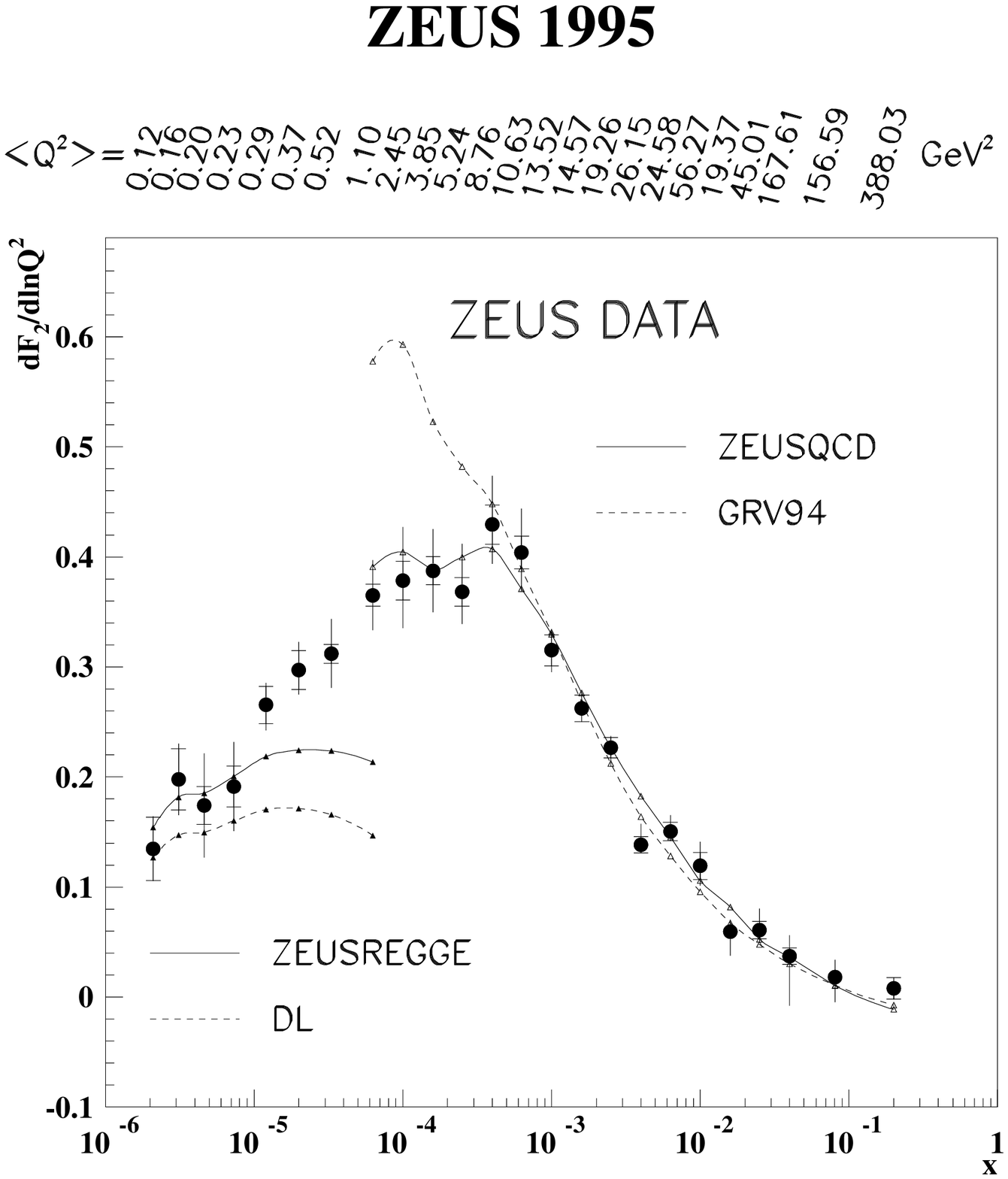,width=7.5cm}
\caption{$dF_2/d\ln Q^2$ versus $x$
and $<\! Q^2\!>$ from fits to ZEUS data of the form
$F_2 = a + b \cdot \ln(Q^2)|_x$.
The data are compared to the QCD and Regge fits discussed in the text.
\label{xn03}}
\end{figure}

To quantify the behaviour of $F_2$, fits to the E665 and ZEUS data 
of the form
$F_2 = c\cdot x^{-\lambda_{\rm eff}}\,|_{Q^2}$ are performed.
The parameter $\lambda_{\rm eff} \simeq \alphapom - 1$ for 
$x~<~0.01$ is 
then plotted as a function of $Q^2$ in Fig.~\ref{xn02}.
A relatively slow transition from $\lambda_{\rm eff} \simeq 0.1$ 
is observed with increasing $Q^2$. 
Also shown are fits to the DL and ZEUSREGGE parameterisations,
fitted over the same $x$ range as the data, for $Q^2 <1$~GeV$^2$.
These describe the data reasonably but are systematically lower.
For $Q^2 >1$~GeV$^2$, the data are compared to the GRV94 prediction
where the starting scale for the evolution of the parton densities is 
rather low ($\sim 0.3$~GeV$^2$) and pQCD evolution generates the rise
at small $x$: this approach is observed to reasonably describe the data 
and the description is improved using the GRV98 PDF (not shown).~\cite{GRV}
The rise at small $x$ is also described by the ZEUSQCD fit, where the
ZEUS data is used as an input.

This rise of $F_2$ with decreasing $x$ is intimately coupled to the
scaling violations via the gluon density 
(in leading order $dF_2/d\ln Q^2 \sim xg(x)$
neglecting sea quark contributions).
In Fig.~\ref{xn03}, fits of the form 
$F_2 = a + b \cdot \ln(Q^2)|_x$ have been performed to the ZEUS 
data and the parameterisations discussed above.
For $x~\sim~2.10^{-4}$, corresponding to $<\! Q^2\! > \sim~4$~GeV$^2$
there is a qualitative change in behaviour where the scaling violations
stabilise and then decrease for lower-$x$ values, a behaviour which is 
not reproduced by the GRV94 PDF.
The question is whether this scaling violation behaviour and the slow onset of
the rise of $F_2$ with decreasing $x$ can be simultaneously
understood. 

A. Mueller has discussed the scaling violation 
behaviour in terms of a geometric model
where the spatial extent, $R_0$, of the $q\bar{q}$ fluctuation of the virtual
photon in the $\gamma^*p$ fixed-target frame 
is related to the height of the plateau in the scaling 
violations~\cite{mueller}
$$\left.\frac{dF_{2}}{d\ln{Q^{2}}}\right|_{x}\simeq
      \frac{Q^{2}(\pi R_{0}^{2})}{4\pi \alpha}.$$
For $\left.\frac{dF_{2}}{d\ln{Q^{2}}}\right|_{x}\simeq 0.4$~GeV$^{-2}$ 
and $Q^2 \simeq 4$~GeV$^2$, the spatial extent $R_0 \simeq 0.3$~fm.
This appears to be the scale at which a transition takes place and 
the partons in the proton start to overlap.~\cite{GLR} 
However, perturbative QCD is pushed to its limit and it will be important to
test that the parton densities extracted from $F_2$ can be universally applied.

The ZEUS NLO fit to the 
$Q^2 > 1$~GeV$^2$ data describes the data, 
demonstrating that there is sufficient flexibility in such an 
approach to go down to relatively low $Q^2$.
However, the relatively stable scaling violations observed 
around $<\! Q^2\! > \sim 4$~GeV$^2$ in 
Fig.~\ref{xn03} yield a gluon contribution which 
is rapidly diminishing
at small-$x$ and which is in fact smaller than the singlet sea quark 
contribution for 
small starting scales.
For larger $Q^2$ values the gluon dominates the sea and 
we have an intuitively appealing picture where gluons radiate sea quarks
whereas, 
in this low-$Q^2$ region, the sea appears to be driving the gluon at low $x$. 
Whether such low-$Q^2$ partons 
are universally valid could be tested using e.g. low-$Q^2$ diffractive
vector meson data.~\cite{martin,mueller}

\begin{figure}
\center
\epsfig{file=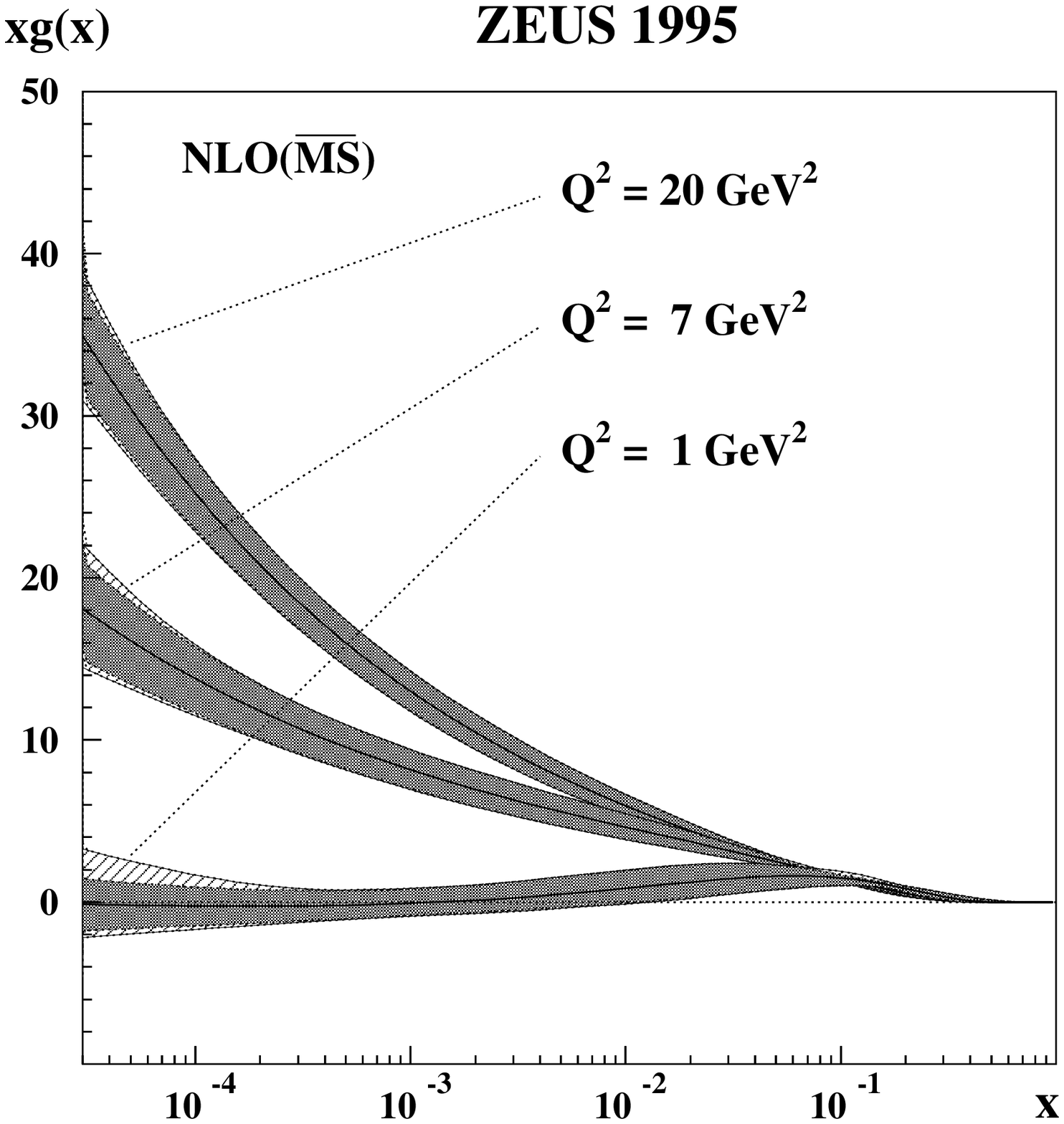,width=10cm}
\caption{Gluon momentum distribution, $xg(x)$ at $Q^2 = 1$, 7 and 20~GeV$^2$
in $\overline{\rm{MS}}$ scheme 
including the uncertainties discussed in the text. 
The shaded band corresponds to the experimental and theoretical uncertainties
and the hatched band indicates the parameterisation errors discussed in
the text.
\label{glse}}
\end{figure}

An important part of the ZEUS NLO QCD fit is the determination of the 
uncertainties on the gluon and singlet quark densities. These are given 
for the gluon distribution in Fig.~\ref{glse}.
The overall uncertainty is estimated by combining in quadrature: 
the experimental systematic
uncertainties on the ZEUS as well as the NMC and BCDMS data; the theoretical
uncertainties on $\alpha_s(M_Z^2)$ by $\pm0.005$, the relative strange quark 
content by $\pm 50\%$ and the charm mass by $\pm 0.2$~GeV;
and, the parameterisation uncertainties on the starting scale,
by varying $1<Q_{\rm o}^2<7$~GeV$^2$,
as well as using a more flexible form of the input in terms of Chebycheff 
polynomials and redefining the $\chi^2$ including $stat.\oplus sys.$
errors.
These variations 
correspond to a precision on
the gluon of $\sim 10\%$ at $Q^2=20$~GeV$^2$
where the renormalisation and factorisation scales are set to 
$\mu_R^2=\mu_F^2=Q^2$. The role of the scale uncertainties is discussed
in~\cite{michiel} with respect to future capabilities to determine 
$\alpha_s(M_Z^2)$ at HERA.
The theoretical and parameterisation uncertainties are
amplified at low $Q^2$ such that the gluon is rather poorly determined 
from the scaling violations of $F_2$ in the transition region discussed 
above.\footnote{
The gluon can even become negative at low $Q^2$ 
(in $\overline{\rm{MS}}$ scheme) but
the physical quantities, $F_2$, $F_2^c$ and $F_L$ remain 
positive within the quoted errors.}
It is clear, however, that the gluon is significantly suppressed at low $Q^2$.

\begin{figure}
\center
\hspace*{-0.5cm}\epsfig{file=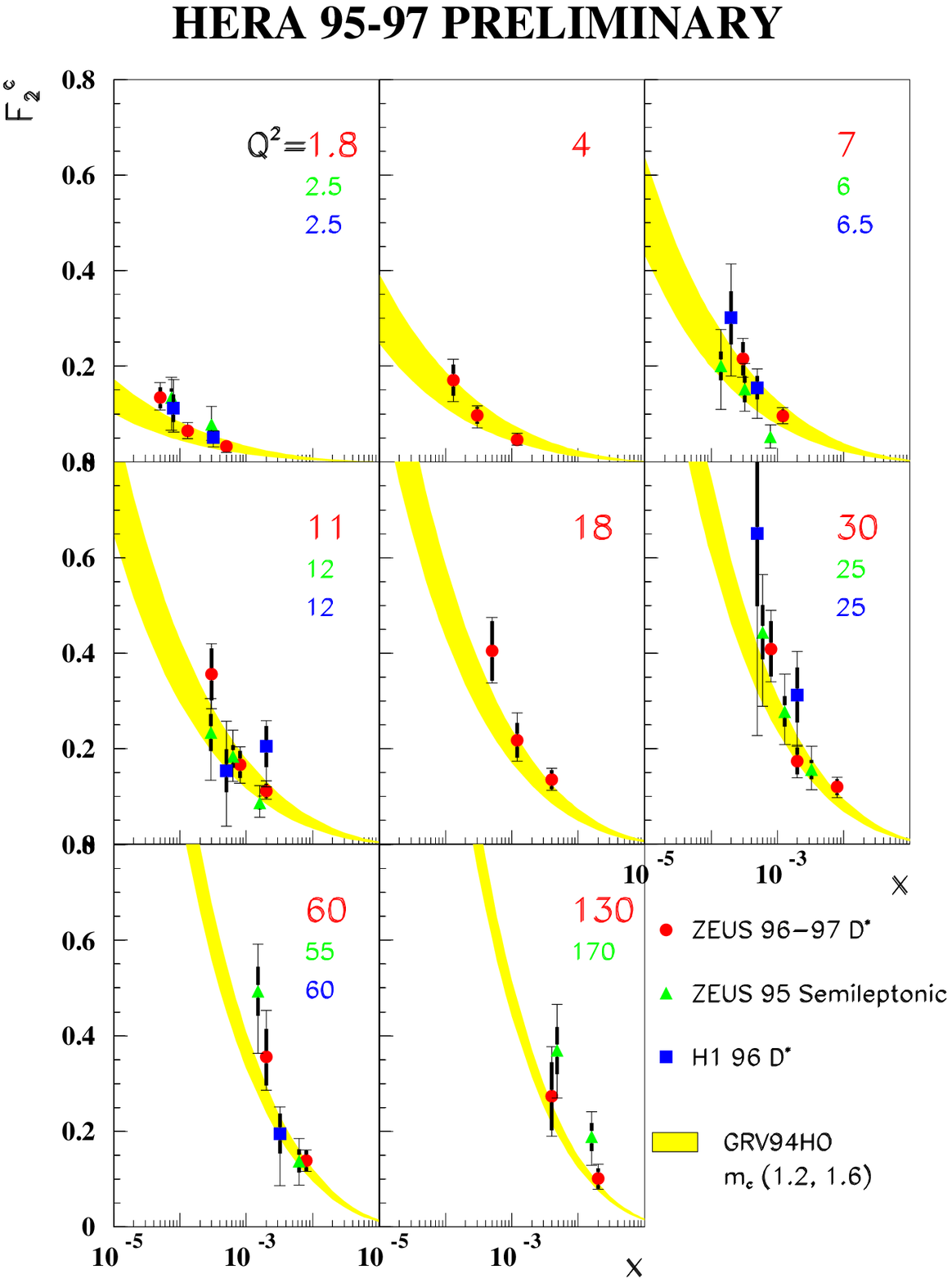,
bbllx=14pt,bblly=22,bburx=384,bbury=564,
width=12cm}
\caption{HERA $F_2^c$ data for various $Q^2$ intervals as a function of
$x$. The data are compared to the Harris-Smith NLO QCD calculation
using the GRV94 PDF input discussed in the text.   
\label{ch02}}
\end{figure}

\noindent{\it $F_2^c$ Determination:}
$D^*\rightarrow (K \pi) \pi_s$ measurements in DIS provide a
significant test of the gluon density of the proton
determined from the scaling violations of $F_2$.
They also help to constrain theoretical uncertainties in the
fits to $F_2$ where different prescriptions for charm production 
are adopted.
The ZEUS preliminary cross-section
$\sigma^{ep\rightarrow D^*X} = 8.55 \pm 0.31 ^{+0.30}_{-0.50}$~nb
is measured in the range
$1<Q^2<600$~GeV$^2$, $0.02<y<0.7,
1.5 < p_T^{D^*} < 15$~GeV, and $|\eta^{D^*}| < 1.5$.
In general, the H1~\cite{katerina} and ZEUS~\cite{wouter} data 
agree with the Harris-Smith NLO calculations~\cite{harris} where
the fraction $f(c\rightarrow D^{*+}) = 0.222\pm0.014\pm0.014$ 
is determined from LEP data~\cite{opal},
the Peterson fragmentation function is characterised by 
$\epsilon_c=0.035$ and the renormalisation and factorisation scales
are set to $\mu_R^2=\mu_F^2=Q^2+4m_c^2$. There is, however, a small discrepancy 
at lower $x_{D^*}$ and higher $\eta^{D^*}$, corresponding to  
the proton direction, where the data lie above the prediction.
A similar discrepancy is also observed in the first analysis in
the semi-leptonic decay channel.~\cite{wouter}
Together these results indicate that the fragmentation of charm
in $ep$ processes is worthy of further investigation.
At this stage, 
it is reasonable to extrapolate the measured cross-section
to the full $\{\eta, p_T \}$ range\footnote{This procedure neglects the
possibility of additional contributions outside the measured region due, for
example, to intrinsic charm.}
to determine $F_2^c(x, Q^2)$ via the expression
$$\frac{d^2\sigma_{c\bar{c}X}}{dx\,dQ^2} =
\frac{2 \pi \alpha^2}{x Q^4}
[ Y_+ F_2^c(x, Q^2) - y^2 F_L^c(x, Q^2)]$$
where the $F_L^c$ contribution can be estimated as a QCD correction.
In Fig.~\ref{ch02}, the HERA $F_2^c(x, Q^2)$ data
mirror the rise of $F_2$ at small $x$.
The data are in agreement with the GRV94 PDF, where the band represents
an estimated theoretical
uncertainty due to the effective charm mass ($m_c = 1.4\pm 0.2$~GeV).
This comparison verifies the steep rise of the gluon
density at low $x$ with a precision of $\simeq 15-20\%$.

\begin{figure}
\center
\hspace*{-0.5cm}\epsfig{file=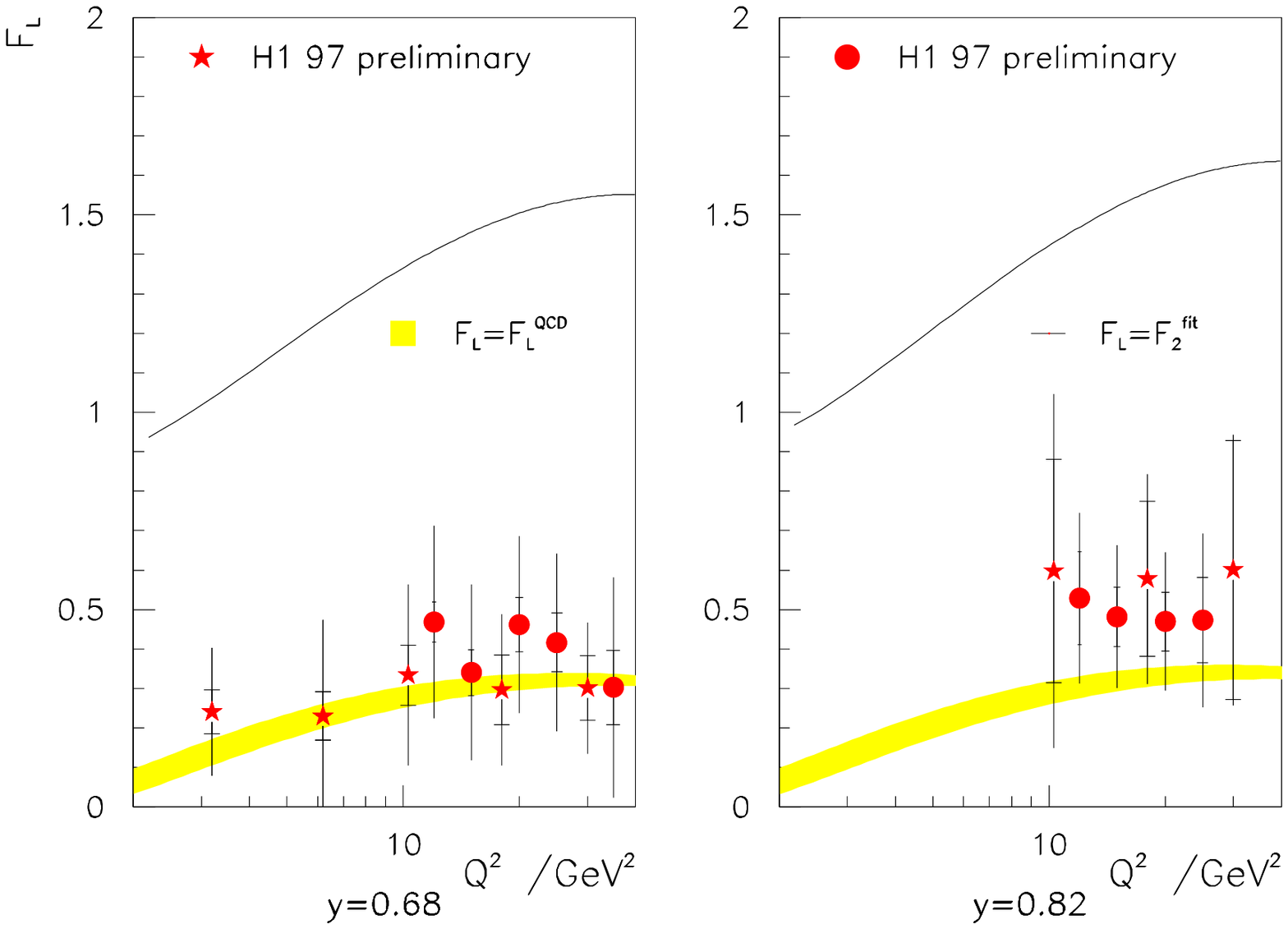,width=15cm}
\caption{H1 determination of $F_L$ versus $Q^2$ for $y=0.68$ and $y=0.82$
using the derivative method ($\star$)
compared to the published method ($\bullet$) and the NLO QCD fit expectation
(shaded band). The upper limit $F_L = F_2$ is indicated by the full line.
\label{defl}}
\end{figure}

\noindent{\it $F_L$ Determination:}
The contribution of $F_L$ enters as a QCD correction to the total DIS 
cross-section where $F_L = F_2 - 2xF_1$. As such it provides an
additional method to calibrate the gluon at low $x$.
H1~\cite{max} have used two methods to extract $F_L$ from the reduced 
cross-section $\tilde{\sigma} = F_2 - \frac{y^2}{Y_+} \cdot F_L$
at high $y$. This is the region where the scattered electron energy
is low: in the H1 analysis scattered positrons are measured down to 
$E_{e'} > 3$~GeV
and backgrounds reduced by requiring the associated track to have correct
charge. $F_L$ is determined as a function of $Q^2 \geq 3$~GeV$^2$
by measuring local derivatives of 
$\partial\tilde{\sigma}/\partial\log y$ and observing deviations 
from a straight line at high $y$.
These data are denoted by the stars in 
Fig.~\ref{defl}. Here, the data are compared to an earlier 
extrapolation method~\cite{H1FL}
applied to the same data,
which yields consistent results, as well as to the H1 NLO QCD fit to H1,
NMC and BCDMS $F_2$ data. The data are in agreement with the QCD expectations
although there is an indication of a relatively large $F_L$ contribution
at the highest $y$ (corresponding to lowest $x$) values.
In conclusion, a consistent value for the gluon density
at low $x \sleq 10^{-2}$ may be
extracted from the data on $F_2$, $F_2^c$ and $F_L$ with a precision of
$\sim 10\%$ at $Q^2=20$~GeV$^2$.

\subsection{High-$Q^2$ Cross-Sections}
The HERA collider provides a unique window to explore $ep$ interactions
at the highest energies, extending the range of momentum transfer $Q^2$
by about two orders of magnitude compared to fixed-target experiments.
As the HERA luminosity increases we explore the region of
$Q^2 \sim 10^4$~GeV$^2$, where electroweak effects play a r\^ole.
It is in this unexplored kinematic region that we are
sensitive to deviations from the standard model (SM).

In 1997, H1~\cite{H1HQ} and ZEUS~\cite{ZEUSHQ} reported an excess of
events compared to the SM predictions from the
neutral current (NC) data taken during 1994 to 1996.
For the H1 analysis an accumulation of 7~events in a reconstructed
$e^+q$ mass window of 200$\pm$12~GeV was found, 
compared to an expectation of $0.95\pm0.18$ from 15~pb$^{-1}$ of data.
One further event was found from the 1997 data 
corresponding to a further 22~pb$^{-1}$, yielding 8~events
compared to an expectation of $3.01\pm0.54$.
For the ZEUS analysis the observed rates agreed with expectations
except for an excess at the highest $Q^2$ where two outstanding events
with $Q^2 \simeq 40,000$~GeV$^2$ were observed from a luminosity of
20~pb$^{-1}$.
These events still clearly stand out but
no new NC outstanding
events are observed in the 1997 data,
corresponding to a further 26.5~pb$^{-1}$.
Similarly in the charged current (CC) channel, the number of events
is higher than expectations but is consistent with the standard model.
Attention has therefore focussed on measuring the cross-sections at the
highest accessible $Q^2$ values.

The theoretical uncertainty on the cross-sections
was determined as discussed w.r.t. the ZEUS low-$Q^2$ NLO QCD fit
using high-$x$ $F_2^p$ and $F_2^n$ data to 
yield SM cross-section uncertainties of
$\simeq$~6-8\% on the NC cross-section and
$\simeq$~6-12\% on the CC cross-section at
the highest accessible $Q^2$ values.
These cross-sections therefore represent a benchmark for the standard model.
The cross-sections, discussed below, are corrected to the electroweak
Born level
and integrated over the measured range of $y$ for H1~\cite{manfred}
and corrected to the complete $y$ range for ZEUS.~\cite{joern}

\noindent{\it Neutral Current Cross-Sections:}
High-$Q^2$ neutral current events are easily identified from the
high-energy scattered positron.
The cross-section is particularly sensitive to the valence $u$-quark
distribution in the proton
$$
\frac{d^2\sigma_{e^+p}}{dx\,dQ^2} \simeq \frac{2 \pi \alpha^2}{x Q^4}
[ Y_+ F_2(x, Q^2) - Y_- xF_3(x, Q^2)].$$
Here, $F_2$ is the generalised structure function,
incorporating $\gamma$ and $Z$ terms, which is sensitive to the singlet
sum of the quark distributions
$(xq + x\overline{q})$ and
$xF_3$ is the parity-violating ($Z$-contribution) term which is
sensitive to the non-singlet difference of the quark distributions
$(xq - x\overline{q})$.
The data are now becoming sensitive to 
electroweak $\gamma Z$ interference effects, 
which suppress the NC cross-section by 
$\sim$ 30\% for $Q^2 > 10,000$~GeV$^2$.~\cite{manfred,joern}

\begin{figure}
\center
\hspace*{-0.5cm}\epsfig{file=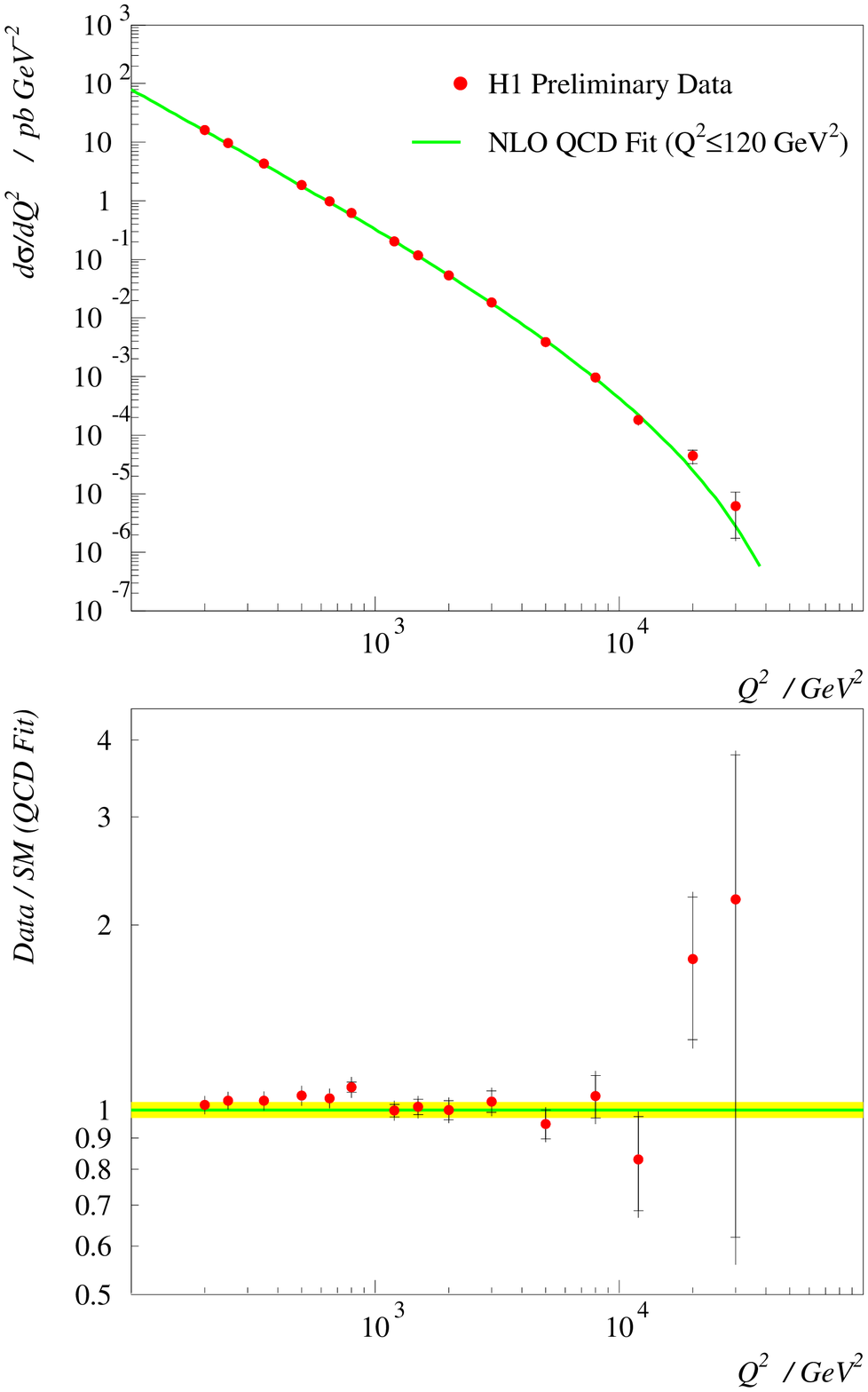,
bbllx=20.,bblly=60.,bburx=540.,bbury=760.,
width=12cm}
\caption{H1 neutral current cross-section
$d\sigma^{NC}/dQ^2$ versus $Q^2$ for $y<0.9$ and $E_{e'} > 11$~GeV 
(upper plot) and ratio
with respect to the standard model prediction (lower plot).
The shaded band represents the luminosity uncertainty of 
$\pm2.6$\%.
\label{ncxs}} 
\end{figure}

In the upper plot of Fig.~\ref{ncxs}, the H1 cross-section is observed
to fall over more than six orders of magnitude.
The ratio of the data to the SM, adopting the H1 NLO QCD fit,
is shown in the lower plot of Fig.~\ref{ncxs} where
agreement is observed up to $Q^2$ $\simeq$~30,000~GeV$^2$.
Comparison of the the data uncertainties
with those from theory (not shown) indicates that the data will
constrain the parton densities of the proton at large $x$.
\begin{figure}
\center
\epsfig{file=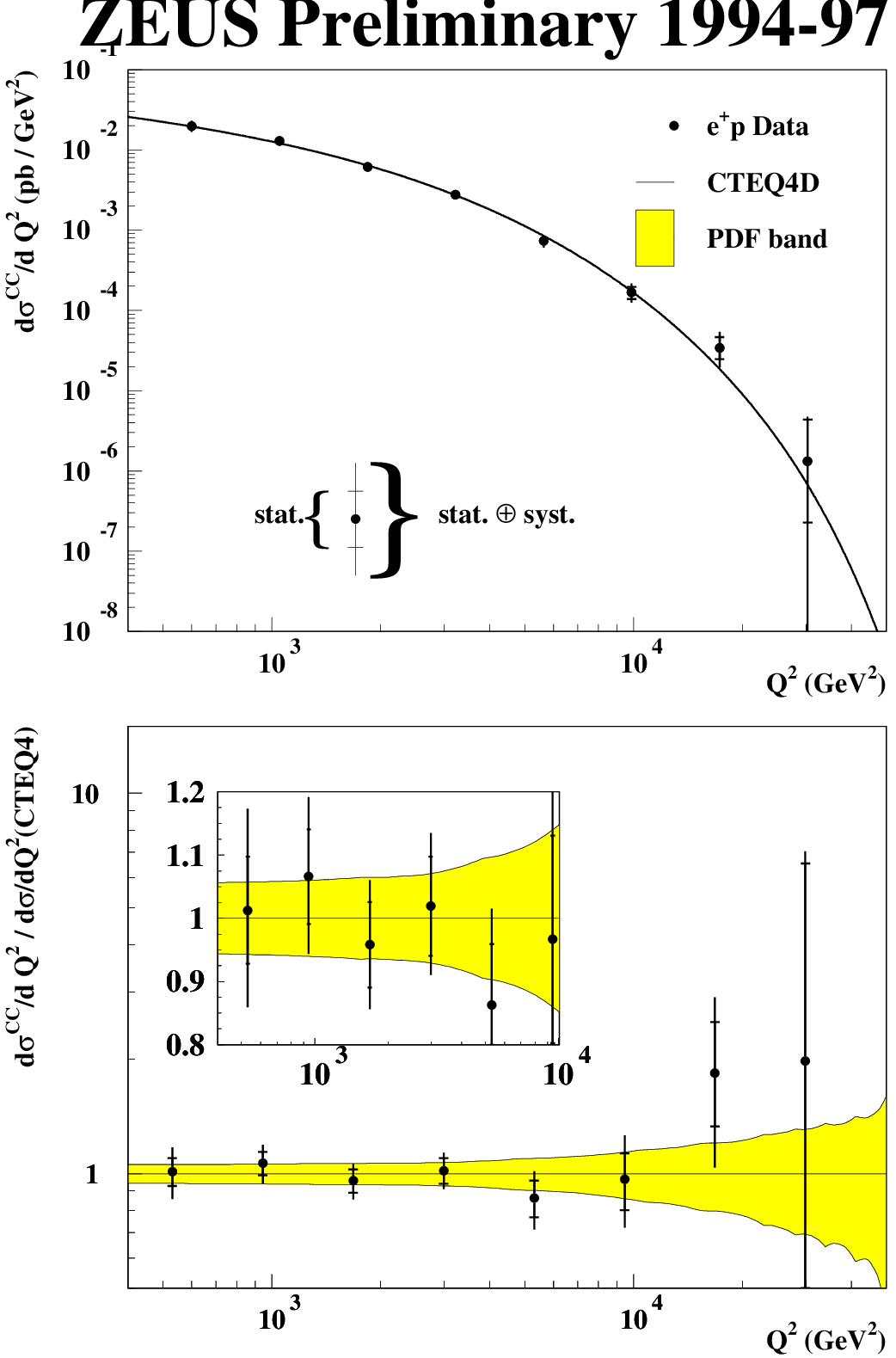,width=10cm}
\caption{ZEUS charged current cross-section 
$d\sigma^{CC}/dQ^2$ versus $Q^2$ for $0<y<1$ 
(upper plot) and ratio 
with respect to the standard model prediction (lower plot).
The shaded band in the ratio plot represents the ZEUS NLO QCD fit uncertainty.
\label{ccxs}}
\end{figure}

\noindent{\it Charged Current Cross-Sections:}
Charged current events are identified by their missing transverse momentum
($p_T$) due to the escaping neutrino.
The cross-section is sensitive to the valence $d$-quark
distribution in the proton
$$\frac{d^2 \sigma_{e^+p}}{dx\, dQ^2} \simeq \frac{G_F^2}{2 \pi}
\left(\frac{M_W^2}{Q^2 + M_W^2}\right)^2 [
\overline{u} + \overline{c} + (1-y)^2 (d + s)]. $$
In the ZEUS analysis, 
$d\sigma^{CC} / dQ^2$ was measured for $Q^2 > 400$~GeV$^2$
using the Jacquet-Blondel method where $Q^2_{JB} = p_T^2/1-y_{JB}$,
with an RMS resolution on $Q^2$ $\simeq 25\%$, reflecting the
$35\%/\sqrt{E}$ hadronic energy resolution.
The systematic uncertainties, mainly due to the hadronic energy scale
uncertainty of $\pm3\%$, correspond to $\sim 15\%$ uncertainties on the
cross-section at lower $Q^2$ but increase at larger $x$ and $Q^2$.

In the upper plot of Fig.~\ref{ccxs} the cross-section is observed
to fall over more than four orders of magnitude.
The ratio of the data to the SM, adopting the CTEQ4D PDF,
is shown in the lower plot of Fig.~\ref{ccxs} where
good agreement is observed up to $Q^2$ of $\simeq$~10,000~GeV$^2$.
Comparison of the the data uncertainties
with those from theory (shaded band) indicates that the data
will help to
constrain the $d$-quark densities at large-$x$.
The cross-section is suppressed 
relative to NC $\gamma$ exchange due to the 
propagator term: this characteristic
dependence on $Q^2$ has been fitted to yield a value for the mass of the
exchanged (space-like) $W$-boson of
\begin{eqnarray}
M_W & = & 78.6 ^{+ 2.5}_{- 2.4}(stat.) ^{+3.3}_{- 3.0}(sys.)\;\rm 
GeV~~~(ZEUS~prelim.) \nonumber
\\
M_W & = & 81.2 \pm3.3 (stat.) \pm4.3(sys.)\;\rm GeV~~(H1~prelim.) \nonumber
\end{eqnarray}
with an additional PDF uncertainty of $\pm 1.5$~GeV estimated in the ZEUS 
analysis and an uncertainty of $\pm 3$~GeV due to electroweak radiative
corrections estimated in the H1 analysis.

\begin{figure}
\center
\epsfig{file=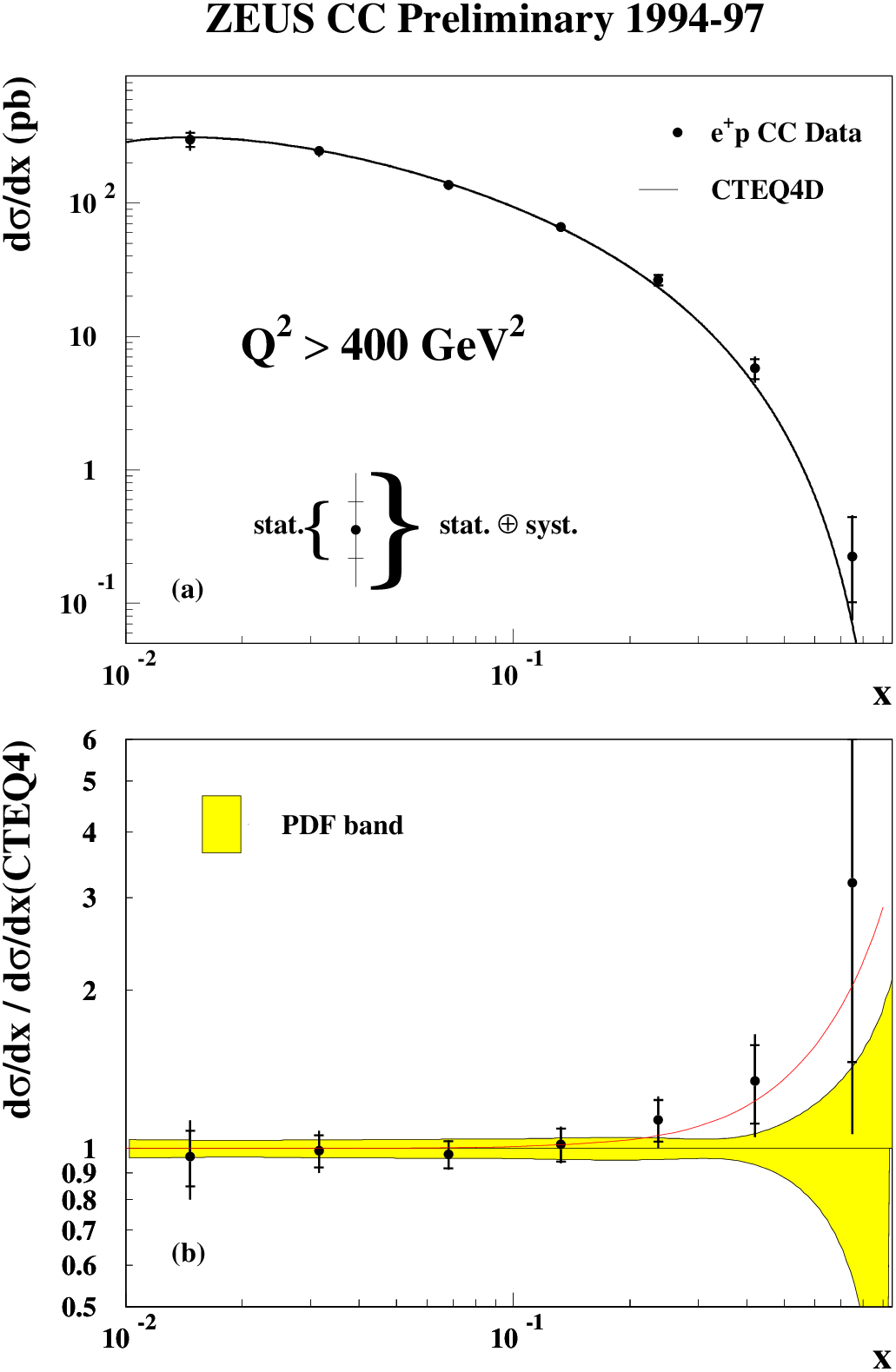,
bbllx=0.,bblly=30.,bburx=610.,bbury=790.,
width=12cm}
\caption{ZEUS charged current cross-section 
$d\sigma^{CC}/dx$ versus $x$ for $Q^2 > 400$~GeV$^2$
and ratio
with respect to the standard model prediction (lower plot).
The upper curve in the ratio plot represents the Bodek-Yang $d/u$ modified
MRS(R2) parameterisation discussed in the text.
\label{ccxx}}
\end{figure}

Returning to the Bodek-Yang analysis,~\cite{yang} discussed in relation to 
the fixed-target results, the ZEUS CC cross-section is plotted as function
of $x$ for $Q^2 > 400$~GeV$^2$ in Fig.~\ref{ccxx}. The increase in the ratio
of $d/u$ corresponds to an increase of the CC cross-section at high $x$.
The uncertainties on the data are large in this region, but this modification
does result in better agreement with the data than the standard PDF's
at large $x$. 

\begin{figure}
\center
\hspace*{-0.3cm}\epsfig{file=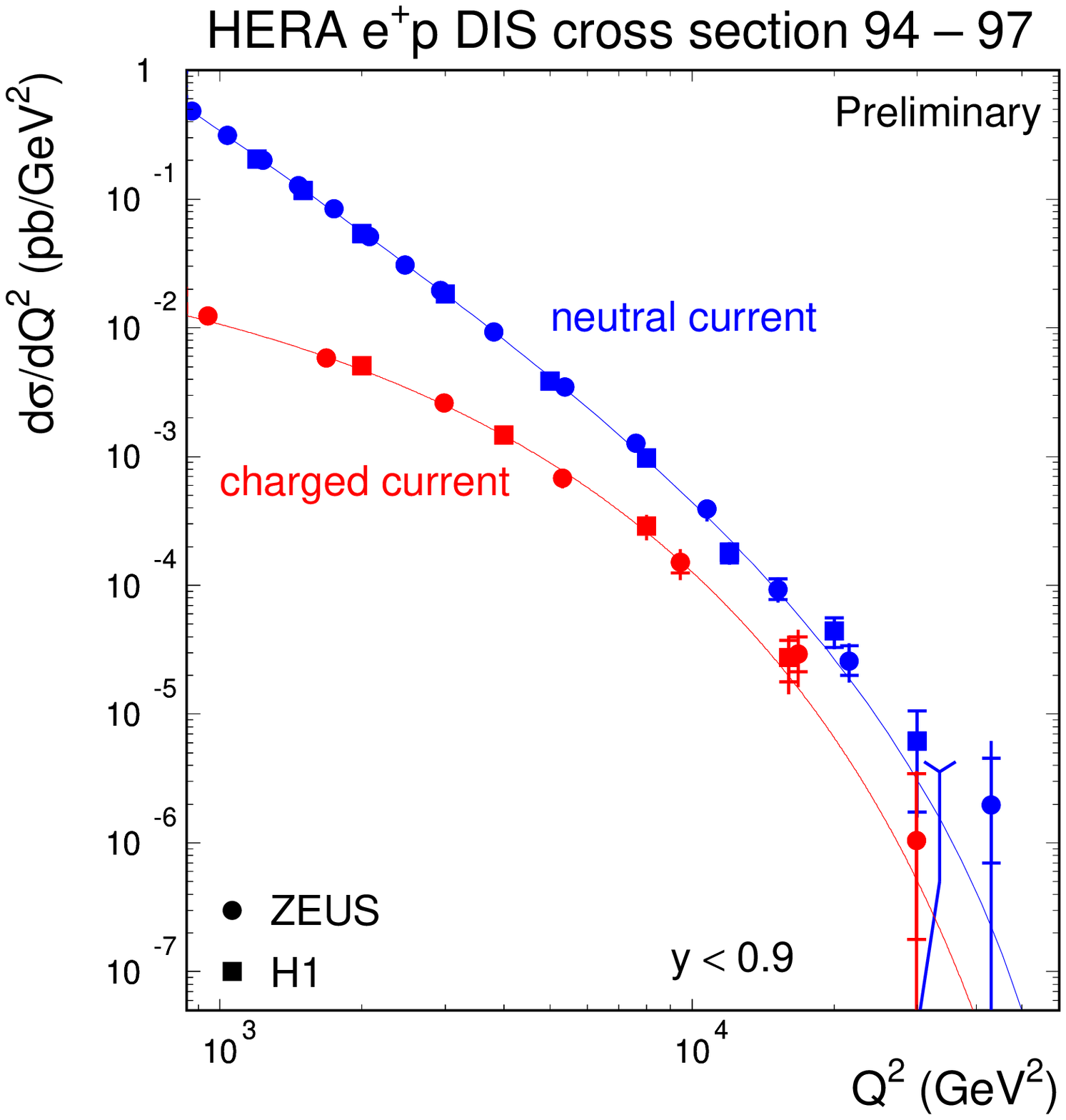,width=12cm}
\caption{HERA $e^+p$ DIS cross-sections 
neutral and charged current data for $y < 0.9$
compared to the standard model prediction, adopting the CTEQ4D PDF.
\label{zhxs}}
\end{figure}

A comparison of the H1 and ZEUS NC and CC cross-sections 
for $Q^2>1000$~GeV$^2$ is given in Fig.~\ref{zhxs}. 
The NC and CC data are in agreement and both cross-sections  
agree with the SM prediction over a broad range range of $Q^2$.
At high $Q^2>10,000$~GeV$^2$, the CC cross-section is suppressed 
relative to the NC cross-section due to the $d/u$ ratio being less than unity.
The measurement of the HERA CC/NC ratio of cross-sections 
will therefore provide a direct determination of this ratio, free from the 
uncertainties due to nuclear binding effects.

\section{Summary and Outlook}
There were many highlights in our deepening understanding of nucleon 
structure presented at the ICHEP98 conference. 
CCFR have performed final analyses of their structure function 
data which lead to precise tests of QCD.
NuTeV already provide electroweak input and aim to reduce the uncertainty
on $\alpha_s(M_Z^2)$ to $\pm0.002$. 
Important developments have been made in the understanding of
higher twist effects in terms of renormalon theory.
Input from various experiments and a reassessment of the importance
of nuclear binding effects in the deuteron lead to the conclusion that
the ratio of $d/u$ parton densities increases, 
whilst the ratio of $\bar{d}/\bar{u}$ decreases
compared to earlier determinations. 

The second phase of spin experiments are now exploring spin structure 
via NLO QCD fits. The first observations
of scaling violations from combined fits to the world data indicate that
the source of the spin puzzle lies with the gluon spin density, $\Delta g$,
although large uncertainties remain. Semi-inclusive measurements now
provide input on the quark composition of the spin.  

The first discovery at HERA was the rise of $F_2$ at low $x$. Precise
data now enable the rise of the associated scaling 
violations with decreasing $x$ and hence the gluon to be determined.
The determinations of $F_2^c$ and $F_L$ enable this gluon distribution 
to be calibrated.
The fall of these scaling violations at low $(x,Q^2)$ enables the region where 
parton confinement effects take place to be explored
at high energies. 

The large $e^+p$ data sample enables cross-sections to be measured at 
very large $Q^2$ where electroweak effects start to play a r\^{o}le.
The HERA data are consistent with the Standard Model and place constraints
on the parton densities at large $x$. The outlook is for a similar sample
of $\sim 50$pb$^{-1}$ of $e^-p$ data in the next two years, prior to
the HERA upgrade where luminosities will be increased five-fold.
Deep inelastic scattering has historically led to the discovery of the 
nucleus, quarks and electroweak neutral currents. 
The discovery potential of current and future experiments is high and
the field continues to provide important input to our understanding 
of sub-nuclear structure.

\section*{Acknowledgements}
It is a pleasure to thank Alan Astbury and co-organisers for 
an excellent conference.
Many thanks to
Halina Abramowicz, 
Arie Bodek,
Antje Brull,
Allen Caldwell, 
Abe Deshpande,
John Dainton,
Robin Devenish,
Martin Erdmann,
Laurent Favart,
Thomas Gehrmann, 
Tim Greenshaw,
Beate Heinemann,
Peppe Iacobucci,
Robert Klanner,
Max Klein, 
Masahiro Kuze,
Ludger Lindemann,
Alan Martin,
Gavin McCance, 
Jan Okrasi\'{n}ski,
Jen-Chieh Peng,
Alex Prinias, 
Robert Waugh,
Arnulf Quadt,
David Saxon,
Stefan Schlenstedt,
Mike Vetterli,
Manuella Vincter,
Bryan Webber,
Jim Whitmore,
Un-Ki Yang,
Rik Yoshida,
Jaehoon Yu 
and all the speakers in the structure functions parallel session
for their help, inspiration and advice.
I am grateful to 
the Alexander von Humboldt Foundation,
DESY and PPARC for their financial support.


\end{document}